\documentclass[%
showpacs,
nofootinbib,
 amsmath,amssymb,
prd,
eqsecnum,
twocolumn
]{revtex4}

\usepackage{graphicx}
\usepackage{dcolumn}
\usepackage{bm}
\usepackage{color}

\newcommand{\be}{\begin{equation}}
\newcommand{\ee}{\end{equation}}
\newcommand{\ben}{\begin{equation}}
\newcommand{\een}{\end{equation}}
\newcommand{\bea}{\setlength\arraycolsep{2pt} \begin{eqnarray}}
\newcommand{\eea}{\end{eqnarray}}
\newcommand{\nnr}{\nonumber \\}

\newcommand{\eq}[1]{(\ref{#1})}

\newcommand{\fr}{\frac}
\newcommand{\tf}{\tfrac}

\newcommand{\wtd}{\widetilde}

\newcommand{\df}{\textrm{d}}
\newcommand{\expe}[1]{\textrm{e}^{#1}}
\newcommand{\pd}{\partial}
\newcommand{\sr}{\sqrt}

\newcommand{\gc}{\gamma}
\newcommand{\gC}{\Gamma}
\newcommand{\gd}{\delta}
\newcommand{\gD}{\Delta}
\newcommand{\gep}{\epsilon}

\newcommand{\gz}{\zeta}
\newcommand{\gq}{\theta}

\newcommand{\gl}{\lambda}

\newcommand{\gvf}{\varphi}

\newcommand{\gf}{\phi}

\newcommand{\im}{\textrm{i}}

\newcommand{\SL}{\textrm{SL}}
\newcommand{\SO}{\textrm{SO}}

\newcommand{\SU}{\textrm{SU}}

\newcommand{\bbR}{\mathbb{R}}

\newcommand{\cA}{\mathcal{A}}
\newcommand{\cF}{\mathcal{F}}
\newcommand{\cL}{\mathcal{L}}
\newcommand{\cM}{\mathcal{M}}
\newcommand{\cN}{\mathcal{N}}

\newcommand{\ds}{\textrm{d} s^2}

\newcommand{\p}{\partial}
\newcommand{\eps}{\epsilon}
\newcommand{\nn}{\nonumber}

\DeclareFontFamily{U}{mathx}{\hyphenchar\font45}
\DeclareFontShape{U}{mathx}{m}{n}{
      <5> <6> <7> <8> <9> <10>
      <10.95> <12> <14.4> <17.28> <20.74> <24.88>
      mathx10
      }{}
\DeclareSymbolFont{mathx}{U}{mathx}{m}{n}
\DeclareMathAccent{\widecheck}{0}{mathx}{"71}

\begin{document}


\title{Dyonic AdS black holes in maximal gauged supergravity}

\author{David D. K. Chow}

\affiliation{%
Physique Th\'eorique et Math\'ematique, \\ Universit\'e Libre de
    Bruxelles and  International Solvay Institutes\\ Campus
    Plaine C.P. 231, B-1050 Bruxelles, Belgium
}

\author{Geoffrey Comp\`ere}
\affiliation{%
Physique Th\'eorique et Math\'ematique, \\ Universit\'e Libre de
    Bruxelles and International Solvay Institutes\\ Campus
    Plaine C.P. 231, B-1050 Bruxelles, Belgium and \vspace{3pt} \\
Center for the Fundamental Laws of Nature, Harvard University,\\
Cambridge, MA 02138, USA
}

\begin{abstract}
We present two new classes of dyonic anti-de Sitter black hole solutions of 4-dimensional maximal $\cN = 8$, $\SO(8)$ gauged supergravity.  They are: (1) static black holes of $\cN = 2$, $\textrm{U}(1)^4$ gauged supergravity with 4 electric and 4 magnetic charges, with spherical, planar or hyperbolic horizons; and (2) rotating black holes of $\cN = 2$, $\textrm{U}(1)^2$ gauged supergravity with 2 electric and 2 magnetic charges.  We study their thermodynamics, and point out that the formulation of a consistent thermodynamics for dyonic anti-de Sitter black holes is dependent on the existence of boundary conditions for the gauge fields. We identify several distinct classes of boundary conditions for gauge fields in $\textrm{U}(1)^4$ supergravity.  We study a general family of metrics containing the rotating solutions, and find Killing--Yano tensors with torsion in two conformal frames, which underlie separability.
\end{abstract}


\maketitle

\tableofcontents


\section{Introduction}


Non-extremal black holes in anti-de Sitter (AdS) spacetime have a crucial role to play in the AdS/CFT correspondence as dual to thermal states of a boundary conformal field theory (CFT) \cite{Maldacena:1997re}. The correspondence is best understood when the gravitational theory is a maximally supersymmetric gauged supergravity theory.  In four dimensions, a gauged theory with maximal number of supercharges exists, namely $\cN = 8$, $\SO(8)$ gauged supergravity \cite{de Wit:1981eq, de Wit:1982ig}\footnote{We are not considering the recently discovered theories \cite{Dall'Agata:2012bb}.}. This theory was originally obtained by gauging a global symmetry of the ungauged $\cN = 8$ supergravity \cite{Cremmer:1978km, Cremmer:1978iv}. Alternatively, it can be obtained by reduction of 11-dimensional supergravity on $S^7$ \cite{de Wit:1986iy}. Notably, the theory admits an $\cN = 2$ abelian truncation to the Cartan subgroup $\textrm{U}(1)^4$ of $\SO(8)$ \cite{Cvetic:1999xp}, and it can be further truncated to $\cN = 2$, $\textrm{U}(1)^2$ supergravity by setting the four gauge fields pairwise equal.

In $\cN = 2$, $\textrm{U}(1)^4$ gauged supergravity, a general black hole solution might carry 4 electric charges and 4 magnetic charges. However, only partial progress has been made in finding a general rotating AdS black hole solution that admits the maximum number of independent charges.  Although general solutions, with the maximum number of independent angular momenta and charges, of analogous theories have been found in $D = 5$ \cite{Wu:2011gq} and $D = 6$ \cite{Chow:2008ip}, less is explicitly known  in $D = 4$.  One reason for further difficulty in $D = 4$ is that a black hole can support both electric and magnetic charges.  Nevertheless, even for the solutions with only electric charges, the gauged generalizations that include rotation are not known.

In ungauged supergravities, the presence of hidden symmetries allows us to generate charged solutions from uncharged solutions using inverse scattering methods or coset methods (see e.g.\ \cite{Belinski:1978aa, Belinski:1979aa, Breitenlohner:1987dg}). In gauged supergravities, the presence of a scalar potential term generically breaks all hidden symmetries, and it is therefore harder to generate black hole solutions of gauged, compared to ungauged, supergravities. More precisely, bosonic Lagrangians of $\cN = 2$ gauged supergravities have the generic form
\ben
\cL_{\textrm{gauged}} = \cL_4  + g^2 V(\Phi^A)\star 1 ,\label{gauged0}
\een
where $g$ is the gauge-coupling constant and $V(\Phi^A)$ is a scalar potential depending upon scalars of the ungauged Lagrangian $\cL_4$. There are a variety of heuristic methods to obtain a new gauged solution by staring at an ungauged solution, guessing the gauged solution, and then checking that the resulting ansatz obeys the gauged supergravity field equations. These methods have been successfully used to obtain several asymptotically AdS black hole solutions of 4-dimensional gauged supergravity \cite{Duff:1999gh, Chong:2004na, Chow:2010sf, Chow:2010fw}, and also in other dimensions.

In this paper we find new static charged solutions of $\cN = 2$, $\textrm{U}(1)^4$ supergravity and rotating charged solutions to $\cN = 2$, $\textrm{U}(1)^2$ supergravity with the maximum number of independent charges. We derive the first law of thermodynamics for these solutions and further discuss a new interplay between boundary conditions and thermodynamics that arises in this context.

We first consider static dyonic black holes of $\textrm{U}(1)^4$ gauged supergravity \cite{Cvetic:1999xp}. We find a 10 parameter family of asymptotically AdS black hole solutions, parameterized by mass, 4 electric charges, 4 magnetic charges, and the gauge-coupling constant, or equivalently the AdS radius, of the theory.  To find this solution, we used the recently found asymptotically flat solution \cite{Chow:2013tia} (see \cite{Cvetic:1995kv} for a more implicit generating solution) of the ungauged theory, which is also known as the STU model (named after the 3 complex scalar fields sometimes denoted S, T and U).  To generalize to a solution of gauged supergravity, it suffices to modify the metric by replacing a single function, with all matter fields unchanged.  The solution generalizes previously known static solutions of this theory \cite{Duff:1999gh, Lu:2013eoa, Lu:2013ura}, which were found by the same method.  This simple method has also been used to find analogous asymptotically AdS solutions in $D = 5$ \cite{Behrndt:1998jd}, $D = 6$ \cite{Cvetic:1999un}, $D = 7$ \cite{Cvetic:1999ne, Liu:1999ai}, and higher dimensions \cite{Chow:2011fh}.

The general solution can, in principle, be embedded into 11-dimensional supergravity. However, explicit formulae are not known in general.  The embedding is explicitly known when there are no axions \cite{Cvetic:1999xp}, which suffices for the solutions discussed in \cite{Duff:1999gh, Lu:2013ura}.  It is also known for when the gauge fields are pairwise equal \cite{Cvetic:1999au}, which suffices for the solutions discussed in \cite{Lu:2013eoa}.

The static solutions just discussed have spherical horizons, but solutions with planar horizons (black branes) are of more interest for studying applications of the AdS/CFT correspondence, since the dual field theory lives on a plane. For example, a particular asymptotically AdS$_4$ electrically charged planar black hole solution of supergravity has been studied in \cite{Gubser:2009qt}, and asymptotically AdS$_4$ dyonic planar black holes with scalars have been studied in \cite{Goldstein:2010aw, Cadoni:2011kv}. Planar black holes are obtained from taking a limit of spherical black holes that effectively zooms in on a narrow cone, say around the north pole. We find the explicit metric and matter fields of the planar black hole with 4 independent electric and 4 independent magnetic charges.  We also analytically continue the solution with a spherical horizon to a solution with a hyperbolic horizon.

We secondly consider rotating dyonic black holes of $\cN = 4$, $\SO (4)$ gauged supergravity \cite{Das:1977pu}.  The theory was originally obtained by gauging a global symmetry of the ungauged $\cN = 4$ supergravity \cite{Das:1977uy, Cremmer:1977tt}.  The explicit reduction ansatz for embedding solutions of $\cN = 4$, $\SO(4)$ gauged supergravity into 11-dimensional supergravity is known \cite{Cvetic:1999au}.  To find black hole solutions, we work with an $\cN = 2$ abelian truncation to $\textrm{U} (1)^2$ gauged supergravity.  This is an $\cN = 2$ gauged supergravity coupled to one vector multiplet.  It is related to the $\textrm{U}(1)^4$ gauged theory by setting the 4 gauge fields pairwise equal.

We find a 7 parameter family of asymptotically AdS black hole solutions, parameterized by mass, rotation, 2 electric charges, 2 magnetic charges, and the gauge-coupling constant, or equivalently the AdS radius, of the theory. We also find the generalization when Newman--Unti--Tamburino (NUT) charge is present, leading to an 8 parameter family of solutions. A special case of the solution is the dyonic Kerr--Newman--Taub--NUT--AdS solution \cite{Plebanski}.  Two special cases have been particularly useful for finding this solution.  One special case is the ungauged limit, which is obtained by setting pairwise equal the 4 gauge fields in the general ungauged solution \cite{Chow:2013tia}.  The second special case is when both gauge fields only contain  electric charges \cite{Chong:2004na}. The gauged solution can be found from the ungauged solution by simply replacing two functions in the metric with everything else untouched, including the matter fields, as done in \cite{Chong:2004na}.  However, whilst this straightforwardly gives a solution locally, more detailed analysis is necessary to examine its physical properties.  In order to write it in an asymptotically AdS coordinate system, an analysis \textit{\`{a} la} Griffiths--Podolsk\'{y} \cite{Griffiths:2005qp} is necessary, which we perform here. It turns out that the metric can be set in a generalized Griffiths--Podolsk\'{y} form with one additional parameter as compared to the non-accelerating metric considered in \cite{Griffiths:2005qp}.

The derivation of the thermodynamic properties of dyonic black holes in AdS requires some care. It turns out that there is a subtle interplay between consistent boundary conditions for the gauge fields and the existence of a consistent thermodynamics. Let us first review the definition of boundary conditions for gauge fields in AdS$_4$. We consider pure Einstein--Maxwell theory for simplicity. Working in radial gauge, $\textrm{U}(1)$ gauge fields admit the Fefferman--Graham expansion
\be
A_i = A^{(0)}_{i}+ \frac{1}{r}A^{(1)}_{i}+O(r^{-2}),
\ee
where $i$ are tangent indices to a surface of constant radius (for definiteness $i=t,\theta,\phi$). The standard Dirichlet boundary condition for the gauge field consists of fixing the boundary magnetic field $b^i = \eps^{ijk}\p_j A^{(0)}_k$. The boundary electric field $e_i = A^{(1)}_{i}$ is then freely varying. These boundary conditions admit the $\SO(3,2)$ conformal group as the asymptotic symmetry group when $b^i = 0$ \cite{Henneaux:1985tv}, and they correspond in the AdS/CFT correspondence to allowing a dual conserved current in the CFT \cite{Witten:1998qj}. Now, an $\SL(2,\mathbb Z)$ ambiguity exists in the quantization of gauge fields in AdS$_4$ \cite{Witten:2003ya}. The S transformation mapping $b^i \rightarrow e_i$, $e_i \rightarrow -b^i$ maps the gauge field $A_\mu$ to a dual gauge field $\widetilde A_\mu$. Applying the standard AdS/CFT correspondence to $\widetilde A_\mu$ is equivalent in terms of the original gauge field $A_\mu$ to imposing Neumann boundary conditions with $e_i$ fixed. The T operation consists of shifting the $\theta$ angle by $2\pi$.  S and T together generate an $\SL(2,\mathbb Z)$ family of boundary conditions. More general Lorentz-invariant mixed boundary conditions can also be imposed, and are dual to conformal field theories with multi-trace deformations \cite{Marolf:2006nd}. For each choice of boundary condition, the electric and magnetic fields cannot be independently varied. This statement corresponds to the fact that the gauge field corresponds to a $\textrm{U}(1)$ current in the boundary CFT and not a $\textrm{U}(1)\times \textrm{U}(1)$ current.

One could then na\"{\i}vely deduce that boundary conditions with both electric and magnetic charges varying in the bulk are generally inconsistent. However, a loophole in the above argument is that only Lorentz-invariant boundary conditions were considered, consistent with the AdS/CFT correspondence. Varying independently electric and magnetic charges simply means that the time components of the electric and magnetic fields $e_{t} = Q$, $b^t = P$ are dynamical. One could still hold fixed the spatial components of the electric and magnetic fields. These boundary conditions are not Lorentz-invariant and the resulting theory will therefore not be dual to a CFT. Note that a dual non-relativistic field theory would also be consistent with the existence of two conserved bulk quantities for a given gauge field. The existence of these boundary conditions therefore has to be studied from first principles in the bulk, independently of any holographically dual picture. One classical criterion for the existence, or not, of these boundary conditions is the existence of a conserved symplectic structure. We performed this analysis which we summarize as follows.

We have four gauge fields $A^I$, with $I = 1, 2,3, 4$, with corresponding electric charges $Q_I$ and magnetic charges $P^I$.  First, the covariant phase space and the mass are defined in the case with only electric charges (all $P^I = 0$) or only magnetic charges (all $Q_I = 0$), which is expected, since these boundary conditions are consistent with a boundary CFT dual description \cite{Witten:2003ya}. Na\"{\i}vely, one would infer the existence of an $\SL(2,\mathbb Z)^4$ family of boundary conditions since there are four gauge fields. However, the gauge fields are coupled to scalar fields, which themselves need to admit consistent boundary conditions. After a detailed analysis, we identified three distinct classes of mixed boundary conditions. The first boundary condition amounts to imposing exactly equal or opposite four electric and magnetic charges $P^I = \pm Q_I$,  $I=1,2,3,4$, with an even number of minus signs. The second boundary condition consists in setting to zero two sets of charges, let say $Q_1=Q_2=P^1=P^2=0$, while imposing on the two remaining sets of charges the constraints $Q_3 = \pm P^3$, $Q_4 = \pm P^4$ for any choice of signs. The third boundary condition consists in setting all but one set of charges to zero, let say $Q_1$ and $P^1$, and impose $Q_1 = \pm P^1$. The last case was discussed in \cite{Lu:2013ura}.  For all such cases, the mass is defined and the first law of thermodynamics holds.

More generally, it turns out that for the generic static black hole with 4 independent electric and 4 independent magnetic charges, the symplectic structure is not conserved. This results in the non-existence of a Hamiltonian, and the black hole mass is ill-defined. We therefore deduce that boundary conditions with such varying fields are not consistent. The non-vanishing symplectic flux at infinity can be seen to be related to the backreaction of the scalar fields due to the gauge fields. The result also holds when only one gauge field is turned on with independent electric and magnetic charges, which explains why the first law does not hold in general in the analysis of \cite{Lu:2013ura}.

Now, there are also cases where Lorentz-violating boundary conditions are consistent (at least in the restricted phase space of black hole solutions). One such example is when the gauge fields are set pairwise equal and both electric and magnetic charges are allowed to be varied. This includes as a subcase the dyonic Reissner-Nordtr\"om--AdS black hole. One can study the first law of thermodynamics using standard treatments and the result is as expected: the first law holds, which provides with a non-trivial check of our expressions.

The rotating solutions that we find possess hidden symmetries in the form of various types of Killing tensors.  This is a widespread feature of charged, rotating black holes in supergravity, in diverse dimensions, that generalize the Kerr solution, see e.g.\ \cite{Chow:2008fe}.  From the form of our rotating solutions, we are motivated to consider a wider class of metrics.  These general metrics possess Killing--Yano tensors with torsion, a known feature of some other black hole solutions in supergravity \cite{Kubiznak:2009qi, Houri:2010fr}.  These are antisymmetric tensors $Y_{\mu \nu}$ satisfying $\nabla_{(\mu}^T Y_{\nu) \rho} = 0$, where the connection has a torsion that is totally antisymmetric.  Unfortunately, for our most general rotating solutions, the physical significance of the torsions is unclear.  In fact, we find these tensors in two different conformal frames, string frame and Einstein frame.  Other types of Killing tensors may be constructed from the Killing--Yano tensors with torsion.  The existence of these Killing tensors is related to the separability of the Hamilton--Jacobi equation for geodesic motion in the two conformal frames, and of the massive Klein--Gordon equation in Einstein frame.

The rest of the paper is organized as follows. We first present in Section \ref{action} two equivalent formulations of the $\cN =2$, $\textrm{U}(1)^4$ gauged supergravity of interest, and review its truncation to $\cN =2$, $\textrm{U}(1)^2$ gauged supergravity. We also provide the formula for its symplectic structure. We present the general static AdS black hole of $\cN =2$, $\textrm{U}(1)^4$ gauged supergravity in Section \ref{BH1} and discuss its thermodynamics. We move to the general rotating black hole of $\cN =2$, $\textrm{U}(1)^2$ gauged supergravity in Section \ref{BH2} and derive its thermodynamics as well.  We generalize the rotating metrics to a wider class of metrics that have various Killing tensors, and study separability.  We finally conclude, and provide an appendix with the details of the general static solution with a spherical horizon.


\section{Gauged supergravities}
\label{action}


The $\cN=2$, $\textrm{U}(1)^4$ gauged supergravity theory is an $\cN = 2$ gauged supergravity coupled to 3 vector multiplets.  The bosonic fields are the metric, 4 $\textrm{U} (1)$ gauge fields $A^I$, 3 dilatons $\gvf_i$ and 3 axions $\chi_i$.  We label the gauge fields by $I = 1, 2, 3, 4$, and label the dilatons and axions by $i = 1, 2, 3$. It is common in the literature to denote $x_i = - \chi_i$ and $y_i = \expe{-\gvf_i}$, which can be united as a complex scalar $z_i = x_i + \im y_i$. Since gauge fields can be dualized in four dimensions, several formulations exist which depend on the duality frame. We will discuss two such formulations. We will also discuss the truncation of $\cN=2$, $\textrm{U}(1)^4$ gauged supergravity to $\cN=2$, $\textrm{U}(1)^2$ gauged supergravity.


\subsection{From maximal gauged supergravity}
\label{sec1}


The original formulation of $\cN=2$, $\textrm{U}(1)^4$ gauged supergravity comes from the $\cN = 2$ abelian truncation of the maximal $\cN = 8$, $\SO (8)$ gauged supergravity that can be obtained from $S^7$ reduction of 11-dimensional supergravity.  The gauged Lagrangian can be found in \cite{Cvetic:1999xp} and consists of the ungauged Lagrangian $\cL_4$ with an extra scalar potential,
\ben
\cL_{\textrm{gauged}} = \cL_4  + g^2 \sum_{i = 1}^3 (2 \cosh \gvf_i + \chi_i^2 \expe{\gvf_i}) \star 1 ,\label{gauged}
\een
where $g$ is the gauge-coupling constant.  We denote the gauge fields in this formulation as $A^I$, for $I=1,2,3,4$. The field strengths are $F^I = \df A^I$ and the dual field strengths are $\wtd{F}_I = \df \wtd{A}_I$, where $\wtd{A}_I$ are dual potentials. The 6 scalar fields have a squared mass equal to $m^2 =- 2g^2$, which lies in the Breitenlohner--Freedman range \cite{Breitenlohner:1982jf},
\be
m_{\textrm{BF}}^2 \leq  m^2 < m_{\textrm{BF}}^2 + g^2 ,
\ee
where $m_{\textrm{BF}}^2=-\tfrac{9}{4}g^2$.

The derivation of the abelian $\cN = 2$ truncation from the full $\cN = 8$, $\SO (8)$ gauged theory treats the 4 gauge fields $A^I$ on an equal footing, corresponding to the $\textrm{U}(1)^4$ Cartan subgroup of the full $\SO(8)$ gauge group \cite{Cvetic:1999xp}. Since dualization is not possible for the general non-abelian gauged theory, this is the preferred formulation of the gauged theory. We shall generally use the terminology ``electric'' and ``magnetic'' according to the nature of $F^I$. This form of the ungauged Lagrangian is complicated, so we shall not use it directly. However, dualization is possible for the abelian truncation, which allows for interesting alternative formulations. We describe one such other formulation in the next section, which has the advantage of a simpler Lagrangian.


\subsection{Dual formulation}
\label{thirdS}


A second formulation of the ungauged theory is from directly reducing 6-dimensional bosonic string theory
\be
\cL_6 = R \star 1 - \tf{1}{2} \star \df \gvf \wedge \df \gvf - \tfrac{1}{2} \expe{-\sqrt{2} \varphi}\star F_{(3)} \wedge F_{(3)}
\ee
on $T^2$, and then dualizing the 4-dimensional 2-form potential $B_{(2)}$ to an axion (Here $F_{(3)}=\df B_{(2)}$).  As before, gauging adds the potential \eqref{gauged}.  The 6-dimensional theory can be regarded as minimal $\cN = 2$ supergravity coupled to a tensor multiplet. After relabelling and changing the signs of some axions, the Lagrangian of \cite{Cvetic:1999xp, Chong:2004na} can be written as
\footnote{Our field strengths are related to the hatted field strengths of \cite{Chong:2004na} by $\cF^1 = \widehat F_{(2){2}}$, $\wtd{\cF}^2 = \widehat F_{(2)1}$, $\wtd{\cF}^3 = \widehat{\cF}_{(2)}^1$, $\cF^4 = \widehat{\cF}_{(2)}^2$ and the signs of $\chi_1$ and $\chi_3$ are flipped while the one of $\chi_2$ is kept fixed. }
\begin{align}
\cL_4 & = R \star 1 - \fr{1}{2} \sum_{i = 1}^3 (\star \df \gvf_i \wedge \df \gvf_i + \expe{2 \gvf_i} \star \df \chi_i \wedge \df \chi_i) \nnr
& \quad - \fr{1}{2} \expe{-\gvf_1} (\expe{\gvf_2 + \gvf_3} \star \cF^1 \wedge \cF^1 + \expe{\gvf_2 - \gvf_3} \star \wtd{\cF}_2 \wedge \wtd{\cF}_2 \nnr
& \quad + \expe{- \gvf_2 + \gvf_3} \star \wtd{\cF}_3 \wedge \wtd{\cF}_3 + \expe{- \gvf_2 - \gvf_3} \star \cF^4 \wedge \cF^4) \nnr
& \quad + \chi_1 (F^1 \wedge F^4 + \wtd{F}_2 \wedge \wtd{F}_3) ,\label{Lthird}
\end{align}
where
\begin{align}
\cF^1 & = F^1 + \chi_3 \wtd{F}_2 + \chi_2 \wtd{F}_3  - \chi_2 \chi_3 F^4 , \nnr
\wtd{\cF}_2 & = \wtd{F}_2 - \chi_2 F^4 , \nnr
\wtd{\cF}_3 & = \wtd{F}_3 - \chi_3 F^4 , \nnr
\cF^4 & = F^4 .
\end{align}
Note that the parity-odd terms may equivalently be written as $\chi_1 (\cF^1 \wedge \cF^4 + \wtd{\cF}_2 \wedge \wtd{\cF}_3)$.

The first formulation is recovered, by dualizing $\wtd{F}_2$ and $\wtd{F}_3$, and by performing the S-duality that replaces $S = \chi_1 + \im \expe{-\gvf_1}$ by $-1/S$.  To dualize $\wtd{F}_2$, we first add to the Lagrangian an extra term
\ben
A^2 \wedge \df \wtd{F}_2 = F^2 \wedge \wtd{F}_2 - \df (A^2 \wedge \wtd{F}_2) .
\een
$A^2$ is a Lagrange multiplier to enforce the Bianchi identity $\df \wtd{F}_2 = 0$.  Varying the modified Lagrangian with respect to $\wtd{F}_2$, we obtain the dual field strength
\begin{align}
F^2 &= \expe{-\gvf_1 + \gvf_2 - \gvf_3} \star (\wtd{\cF}_2 +\chi_3 e^{2\varphi_3} \cF^1) \nnr
& \quad - \chi_1 (\wtd{\cF}_3+\chi_3 F^4) .
\label{dualF21}
\end{align}
One can obtain similarly the expression for $F^3$. Subtituting $\wtd{F}_2$ and $\wtd{F}_3$ in terms of $F^2$ and $F^3$ and performing the S-duality leads to the formulation of the previous section in terms of $(A^1, A^2, A^3, A^4)$.

It is also useful to write the Lagrangian \eqref{Lthird} in the general form
\begin{align}
	\cL_4 & = \df^4 x \, \sqrt{-g} [ R - \tfrac{1}{2}f_{AB}(\Phi)
	\partial_\mu \Phi^A \partial^\mu \Phi^B  \nnr
& \quad - \tfrac{1}{4}k_{IJ}(\Phi)\mathbf F^I_{\mu\nu}\mathbf F^{J\mu\nu}
	+ \tfrac{1}{4}h_{IJ}(\Phi)\epsilon^{\mu\nu\rho\sigma}\mathbf F^I_{\mu\nu} \mathbf F^J_{\rho\sigma} ] , \nnr
\label{generalaction}
\end{align}
where $\Phi^A = (\varphi_1,\varphi_2,\varphi_3,\chi_1,\chi_2,\chi_3)$ are the scalar fields and $\mathbf A^I = (A^1,\widetilde A_2,\widetilde A_3,A^4)$ are the $\textrm{U}(1)$ gauge fields. The kinetic coefficients $f_{AB}$, $h_{IJ}$ are
\begin{align}
f_{AB}& = \textrm{diag} (1,1,1,\expe{2\varphi_1},\expe{2\varphi_2},\expe{2\varphi_3}), \nnr
h_{IJ}& = - \fr{\chi_1}{2}
\begin{pmatrix}
 0 & 0 & 0 & 1  \\
 0 & 0 & 1  & 0 \\
 0 & 1  & 0 & 0 \\
 1 & 0 & 0 & 0
\end{pmatrix} ,
\end{align}
and $k_{IJ}$ is a longer expression that can be easily deduced from \eqref{Lthird}.


\subsection{Truncation to $\textrm{U}(1)^2$ gauged supergravity}


We can consistently truncate the second formulation of $\textrm{U}(1)^4$ gauged supergravity in \eq{Lthird} by setting $A^1 = A^4$, $\wtd{A}_2 = \wtd{A}_3$, $\gvf_2 = \gvf_3 = \chi_2 = \chi_3 = 0$, obtaining
\begin{align}
\cL_4 & = R \star 1 - \tf{1}{2} \star \df \gvf \wedge \df \gvf - \tf{1}{2} \expe{2 \gvf} \star \df \chi \wedge \df \chi \nnr
& \quad - \expe{-\gvf} (\star F^1 \wedge F^1 + \star \wtd{F}_2 \wedge \wtd{F}_2) + \chi (F^1 \wedge F^1 \nonumber \\
& \quad + \wtd{F}_2 \wedge \wtd{F}_2) + g^2 (4 + \expe{\gvf} + \expe{- \gvf} + \chi^2 \expe{\gvf}) \star 1 , \label{LU2}
\end{align}
where $\varphi \equiv \varphi_1$, $\chi \equiv \chi_1$.
Dualizing $\wtd{F}_2$ as in \eq{dualF21}, we have
\be
F^2 = \expe{-\gvf} \star \wtd{F}_2 - \chi \wtd{F}_2 .
\ee
We then obtain the Lagrangian
\begin{align}
\cL_4 & = R \star 1 - \tf{1}{2} \star \df \gvf \wedge \df \gvf - \tf{1}{2} \expe{2 \gvf} \star \df \chi \wedge \df \chi \nnr
& \quad - \expe{-\gvf} \star F^1 \wedge F^1 + \chi F^1 \wedge F^1 \nonumber \\
& \quad - \fr{1}{1 + \chi^2 \expe{2 \gvf}} (\expe{\gvf} \star F^2 \wedge F^2 + \chi \expe{2 \gvf} F^2 \wedge F^2) \nnr
& \quad + g^2 (4 + \expe{\gvf} + \expe{- \gvf} + \chi^2 \expe{\gvf}) \star 1 ,\label{SO4}
\end{align}
which is a bosonic abelian truncation of $\cN = 4$, $\SO (4)$ gauged supergravity.

If we further consistently truncate to $A \equiv A^1 = A^2$ and $\gvf = \chi = 0$, we obtain
\be
\cL_4 = R \star 1 - 2 \star F \wedge F + 6 g^2 \star 1 ,
\ee
which is simply Einstein--Maxwell theory with a negative cosmological constant. This is the bosonic sector of $\cN = 2$ minimal gauged supergravity.


\subsection{Symplectic structure}


One fundamental object of interest of any theory is the symplectic structure, which is a two-form on the space spanned by all the fields, which we denote as $\{\Phi^i\}$. The covariant phase space definition (see e.g.\ \cite{Iyer:1994ys}) is
\be
\bm{\omega}(\delta_1 \Phi^i , \delta_2 \Phi^i) = \delta_2 \bm{\Theta} (\delta_1 \Phi^i) - \delta_1  \bm{\Theta} (\delta_2 \Phi^i) , \label{sympl}
\ee
where $ \bm{\Theta}$ is the pre-symplectic form defined from the variation of the action
\be
\delta \bm{L} = \frac{\delta \bm{L}}{\delta \Phi_i}\delta \Phi_i + \textbf{d}\bm{\Theta} (\delta \Phi) .
\ee
The invariant symplectic structure defined solely from the equations of motion \cite{Barnich:2007bf} differs from the covariant phase space definition by a boundary term. Since this boundary term has no influence on the present discussion, we will simply ignore it and only present the covariant phase space definition.

The pre-symplectic form for a Lagrangian of the form $\cL_4/16 \pi G$, with $\cL_4$ given by the general form \eqref{generalaction}, can be worked out. The fields are $\Phi^i = (g_{\mu\nu},\Phi^A,\mathbf A^I)$. The result is (see e.g.\ \cite{Compere:2009dp})
\begin{align}
\bm{\Theta} & = \star \bm{X} = \frac{1}{3!}\eps_{\mu\alpha_1\alpha_2\alpha_3} X^\mu \, \df x^{\alpha_1}\wedge \df x^{\alpha_2}\wedge \df x^{\alpha_3} \ , \nnr
  X^\mu & = \frac{1}{16\pi G}\big[
(\nabla_\nu h^{\nu\mu} - \nabla^\mu h) - \,f_{AB}(\Phi)\nabla^\mu\chi^B \delta\chi^A \nnr
& \quad  - \, (k_{IJ}(\Phi)\mathbf F^{J\mu\nu}
 - \, h_{IJ}(\Phi)\epsilon^{\mu\nu\rho\sigma}\mathbf F^J_{\rho\sigma})\delta \mathbf A_\nu^I \big].
\end{align}
Here we defined $h_{\mu\nu}=\delta g_{\mu\nu}$, $h^{\mu\nu} = g^{\mu\alpha}g^{\nu\beta}h_{\alpha\beta}$ and $h = g^{\alpha\beta}h_{\alpha\beta}$.


\section{$\textrm{U}(1)^4$ dyonic static black holes}
\label{BH1}


A generating solution for the general asymptotically flat, static, non-extremal, dyonic black hole of ungauged $\cN = 2$, $\textrm{U}(1)^4$ supergravity (and, more generally, for $\cN = 8$ supergravity), also known as the STU model, was obtained a long time ago \cite{Cvetic:1995kv}.  More recently, the solution with 8 explicit electromagnetic charges was given in a simplified form in \cite{Chow:2013tia}.  The spherically symmetric metric has the form
\be
\ds = -\frac{R}{W} \, \df t^2 + \frac{W}{R} \, \df r^2  + W (\df \theta^2 + \sin^2 \theta \, \df \phi^2) , \label{metric1s}
\ee
where $R(r)$ is a quadratic polynomial in $r$, and $W^2(r)$ is a quartic polynomial in $r$. It turns out that it suffices to modify the metric by replacing $R$ by $R+g^2 W^2$ and one obtains a solution of $\cN = 2$, $\textrm{U}(1)^4$ gauged supergravity with potential \eqref{gauged}, with all matter fields unchanged.  Indeed, we checked that the metric accompanied with the matter fields obeys the field equations using Mathematica \cite{Mathematica}. Static AdS$_4$ black holes have been found by this method previously \cite{Duff:1999gh, Lu:2013eoa, Lu:2013ura}, but the function $W^2$ is substantially more complicated in the general case, because it does not factorize.  The solution depends upon 10 arbitrary parameters: the mass parameter $m$, the gauge-coupling constant $g$, four electric charge parameters ($\delta_I$), and four magnetic charge parameters ($\gamma_I$), for $I=1,2,3,4$.  Regular supersymmetric static black holes with spherical or planar horizons do not exist in the purely electric or purely magnetic cases \cite{Duff:1999gh}. Regular supersymmetric static black holes do however exist when the horizon is hyperbolic \cite{Caldarelli:1998hg}.

Let us first present the static black holes with spherical horizons.  We will then present the limit with planar horizons, which presents interesting simplifications.  Static solutions with hyperbolic horizons can be obtained by analytic continuation of the static solutions with spherical horizons.


\subsection{Spherical black hole}



\subsubsection{Metric}


The metric is
\begin{align}
\df s^2 & = -\frac{R + g^2 W^2}{W} \, \df t^2 + \frac{ W \, \df r^2}{R + g^2 W^2} \nnr
& \quad + W ( \df \theta^2 + \sin^2 \theta \, \df \phi^2 ) , \label{metric1}
\end{align}
where
\begin{align}
W^2(r) & = R^2(r)+2 R(r) (2M r+V )+L(r)^2,\label{defW2}\\
R(r) & = r^2 -2 m r -n_0^2,\label{defR} \\
L(r) & = \lambda_1 r + \lambda_0 .
\end{align}
Here we already note that $M$ is the physical mass, when it can be defined, which we will discuss in Section \ref{thermo}. The five constants $M,n_0, \lambda_1, \lambda_0 ,V$ are functions of the parameters $m,\gamma_I,\delta_I$ and can be expressed as follows,
\begin{align}
\label{mass}
M & = m \mu_1 + n_0 \mu_2 , \\
n_0 & = - m \frac{\nu_1}{\nu_2},\label{n0d}\\
\lambda_1 & = 2(m \nu_2 - n_0 \nu_1), \\
\lambda_0 & = 4(m^2 +n_0^2)D, \\
V & = 2 (n_0 \mu_1 - m \mu_2)n_0 +2 (m^2+n_0^2) C,
\end{align}
where the coefficients $\mu_1,\mu_2,\nu_1,\nu_2,D,C$ are functions of only the charge parameters $(\delta_I,\gamma_I)$, and are given in \eq{munu}, \eq{Dconstant} and \eq{Cconstant}. We choose the orientation $\varepsilon_{tr\theta \phi}=1$.

The metric \eqref{metric1} in the ungauged case $g=0$ reduces to the one presented in \cite{Chow:2013tia} upon setting the NUT charge to zero and  identifying $V(n_0)$ in \cite{Chow:2013tia} with $V$ here. The parametrization is asymmetrical between the electric ($\delta_I$) and magnetic ($\gamma_I$) parameters but symmetrical under the exchange of indices $I = 1,2,3,4$. This asymmetry is rooted in the way the solution was obtained in \cite{Chow:2013tia} from coset model techniques, by first magnetic charging and then electric charging the Kerr--Taub--NUT seed with original mass and NUT charges $(m,n)$. The total NUT charge can be cancelled at the end of the process by fixing the NUT parameter $n = n_0(m,\delta_I,\gamma_I)$ in terms of the remaining parameters, which results in the explicit expression for $n_0$ displayed in \eqref{n0d}. 
While it would be interesting to obtain a symmetric parametrization of the black hole, the general features of the solution (the radial dependence of the various functions and all physical properties) do not depend upon the parametrization.


\subsubsection{Matter}


The gauge fields and dual gauge fields are
\begin{align}
A^I & = \gz^I \, \df t + P^I \cos \gq \, \df \phi , & \wtd{A}_I & = \wtd\gz_I \, \df t - Q_I \cos \gq \, \df \phi ,
\end{align}
where the electromagnetic scalars $\zeta^I$ and $\wtd \zeta_I$ are
\begin{align}
\zeta^I &= \frac{- L (P^I n_0 + L^I) + R (Q_I r + V^I )}{W^2} , \nnr
\wtd\zeta_I &= \frac{L (Q_I n_0 +\widetilde L_I) + R (P^I r + \widetilde V_I)}{W^2} .
\end{align}
The electric charges $Q_I$ and magnetic charges $P^I$ are
\begin{align}
Q_I &= 2 \bigg( m \frac{\p \mu_1}{\p \delta_I}+ n_0 \frac{\p \mu_2}{\p \delta_I} \bigg) \equiv m \rho_I^1 + n_0 \rho_I^2,\nnr
P^I &= -2 \bigg( m\frac{\p \nu_1}{\p \delta_I} + n_0 \frac{\p \nu_2}{\p \delta_I} \bigg) \equiv m \pi^I_1 + n_0 \pi^I_2,
\end{align}
where the last expressions are the definitions of $\rho_I^1,\,\rho_I^2,\,\pi^I_1,\,\pi^I_2$ in terms of $(\gd_I, \gc_I)$.  The remaining functions $L^I(r)$ and $\widetilde L_I(r)$ are given in \eq{linearfunctions}, and the constants $V^I$ and $\widetilde V_I$ are given in \eq{Vconstants}.

The scalar fields are of the form
\begin{align}
\chi_i & = \frac{f_i}{r^2 + n_0^2 + g_i} , & \expe{\gvf_i} & = \frac{r^2 + n_0^2 + g_i}{W} ,\label{defsc}
\end{align}
where
\begin{align}
f_{i} & = 2 (m r + n_0^2) \xi_{i 1} - 2 n_0 (r -  m) \xi_{i 2} + 4 (m^2 + n_0^2) \xi_{i 3} , \nnr
g_{i} & = 2 (m r + n_0^2) \eta_{i 1} -  2 n_0 (r  -  m) \eta_{i 2} + 4 (m^2 + n_0^2) \eta_{i 3} ,
\end{align}
and $\xi_{i1},\,\xi_{i2},\,\xi_{i3},\,\eta_{i1},\,\eta_{i2},\,\eta_{i3}$ are given in \eq{xieta} as functions of the electromagnetic parameters $(\delta_I,\gamma_I)$.


\subsubsection{Special cases}


If $\gd_I = \gd$ and $\gc_I = \gc$ for $I=1,2,3,4$, then we have the dyonic Reissner--Nordstr\"{o}m--AdS solution of Einstein--Maxwell theory.  The conserved charges are then
\begin{align}
M & = \fr{m [1 + \cosh (4 \gd) \cosh (4 \gc)]}{2 \cosh (2 \gd) \cosh (2 \gc)} , \nnr
Q_I & = \fr{m \sinh (2 \gd) \cosh (4 \gc)}{\cosh (2 \gc)} , \nnr
P^I & = \fr{m \sinh (2 \gc)}{\cosh (2 \gd)} ,
\end{align}
for each $I$.  If we define $r' = r + M - m$, $Q = Q_I$ and $P = P^I$, then the metric is
\be
\df s^2 = - f \, \df t^2 + \fr{\df r'{^2}}{f} + r'{^2} (\df \gq^2 + \sin^2 \gq \, \df \phi^2) ,
\ee
where $f = 1 - 2 M/r' + (Q^2 + P^2)/r'{^2} + g^2 r'{^4}$, and the gauge fields are
\begin{align}
A^I & = \fr{Q}{r'} \, \df t + P \cos \gq \, \df \phi , & \wtd{A}_I & = \fr{P}{r'} \, \df t - Q \cos \gq \, \df \phi ,
\end{align}
which is recognizable as dyonic Reissner--Nordstr\"{o}m--AdS.

If $g = 0$, then we have the static and asymptotically flat limit of the general ungauged solution, which has recently been found \cite{Chow:2013tia}.  It is a solution of the so-called STU model.

If $\gc_I = 0$, then we have the static AdS black hole with 4 electric charges \cite{Duff:1999gh}.  If $\gd_I = 0$, then we have the static AdS black hole with 4 magnetic charges \cite{Duff:1999gh}.

If $\gd_2 = \gd_3 = \gd_4 = \gc_2 = \gc_3 = \gc_4 = 0$, then we have the static AdS black hole with 1 dyonic gauge field \cite{Lu:2013ura}.

If $\gd_1 = \gd_2$, $\gc_1 = \gc_2$ and $\gd_3 = \gd_4 = \gc_3 = \gc_4 = 0$, then we have the static AdS black hole with 2 equal dyonic gauge fields \cite{Lu:2013eoa}.


\subsection{Planar black hole}


Static black holes with planar horizons and 8 independent electromagnetic charges can be obtained as a limit of the solution with spherical horizons. We replace $t \rightarrow \gep t$, $r \rightarrow r/\gep$, $\gq \rightarrow \gep \rho$, $m \rightarrow m/\gep^3$, $\gd_I \rightarrow \gep \gd_I$, $\gc_I \rightarrow \gep \gc_I$, and then take the limit $\gep \rightarrow 0$.  We can then switch from plane polar coordinates $(\rho, \phi)$ to cartesian coordinates $(x, y)$.  We use the notation $\delta_{IJ}=\delta_I \delta_J$, $\delta_{1234}=\delta_1 \delta_2\delta_3 \delta_4$, $\gamma_{IJ}=\gamma_I \gamma_J$, $\gamma_{1234}=\gamma_1 \gamma_2\gamma_3 \gamma_4$.

The solution with planar horizons has the metric
\be
\ds = -f \, \df t^2 +\frac{\df r^2}{f} + W ( \df x^2 + \df y^2 ) ,
\ee
where
\begin{align}
f(r) &= \frac{- 2 m r + g^2 W^2}{W}, \nnr
W^2 (r) & = r^4 + w_3 m r^3 + w_2 m^2 r^2 + w_1 m^3 r + w_0 m^4 ,
\end{align}
and where
\begin{align}
w_3 & = 2 \textstyle \sum_I (\gd_I^2 + \gc_I^2) , \nnr
w_2 & = 2 [ 3 \textstyle \sum_I \gd_I^2 \gc_I^2 + 2 \textstyle \sum_{I < J} ( \gd_{I J}^2 + \gc_{I J}^2 - \gd_{I J} \gc_{I J} \nnr
& \quad + 2 \gc_{1 2 3 4} \gd_{I J}/\gc_{I J} ) ] , \nnr
w_1 & = 2 \{ ( \textstyle \sum_I \gd_I \gc_I )^2 \sum_J (\gd_J^2 + \gc_J^2) + 4 \textstyle \sum_I ( \gd_{1 2 3 4}^2/\gd_I^2 \nnr
& \quad + \gc_{1 2 3 4}^2/\gc_I^2 - \gd_{1 2 3 4} \gc_I^2 - \gc_{1 2 3 4} \gd_I^2 ) \nnr
& \quad + 4 \textstyle \sum_{I < J} [ \gd_{1 2 3 4} \gc_{I J} ( \gd_I/\gd_J + \gd_J/\gd_I ) \nnr
& \quad + \gc_{1 2 3 4} \gd_{I J} ( \gc_I/\gc_J + \gc_J/\gc_I ) - \gd_{I J} \gc_{I J} (\gd_I^2 + \gd_J^2 + \gc_I^2 \nnr
& \quad + \gc_J^2) ] \} , \nnr
w_0 & = [ 4 (\gd_{1 2 3 4} + \gc_{1 2 3 4}) - \textstyle \sum_I \gd_I^2 \gc_I^2 + 2 \sum_{I < J} \gd_{I J} \gc_{I J} ] ^2 .
\end{align}
Note that the coefficient $w_0$ is the square of a Cayley hyperdeterminant, and is invariant under $\SL(2, \bbR)^3$ transformations, see e.g.\ \cite{Duff:2006uz}.  We choose the orientation $\varepsilon_{t r x y} = 1$.

The gauge fields and dual gauge fields are
\begin{align}
A^I & = \gz^I \, \df t - \tf{1}{2} P^I  (x \, \df y - y \, \df x ), \nnr
\wtd{A}_I & = \wtd\gz_I \, \df t + \tf{1}{2} Q_I  \,  (x \, \df y - y \, \df x ) .
\end{align}
The electromagnetic charges are
\begin{align}
Q_I & = 2 m \gd_I , & P^I & = 2 m \gc_I .
\end{align}
The electromagnetic scalars can be expressed concisely as
\begin{align}
\gz^I & = \fr{1}{W^2} \bigg( \fr{1}{2} \fr{\pd (W^2)}{\pd \gd_I} + 2 m^3 r \gc_I Z \bigg) ,\nnr
\wtd\gz_I & = \fr{1}{W^2} \bigg( \fr{1}{2} \fr{\pd (W^2)}{\pd \gc_I} - 2 m^3 r \gd_I Z \bigg) ,
\end{align}
where
\begin{align}
Z & = 2 \sum_I \bigg( (\gc_I^2 - \gd_I^2) \gd_I \gc_I + \gc_{1 2 3 4} \fr{\gd_I}{\gc_I} - \gd_{1 2 3 4} \fr{\gc_I}{\gd_I} \bigg) \nnr
& \quad + \sum_I  (\gd_I^2 - \gc_I^2) \sum_J \gd_J \gc_J .
\end{align}

The scalars are
\begin{align}
\chi_1 & = \fr{2 m [r (\gd_2 \gc_3 + \gd_3 \gc_2 - \gd_1 \gc_4 - \gd_4 \gc_1) + m f_1]}{r^2 + 2 m r (\gd_2^2 + \gd_3^2 + \gc_1^2 + \gc_4^2) + m^2 g_1} , \nnr
\expe{\gvf_1} & = \fr{r^2 + 2 m r (\gd_2^2 + \gd_3^2 + \gc_1^2 + \gc_4^2) + m^2 g_1}{W} ,
\end{align}
where
\begin{align}
f_1 & = (\gd_2 \gc_2 + \gd_3 \gc_3 - \gd_1 \gc_1 - \gd_4 \gc_4 ) (\gd_{1 4} + \gd_{2 3} + \gc_{1 4} + \gc_{2 3}),\nnr
g_1 & = 4 (\gd_{2 3} + \gc_{1 4})^2 + (\gd_1 \gc_1 + \gd_4 \gc_4 - \gd_2 \gc_2 - \gd_3 \gc_3)^2,
\end{align}
with $\varphi_2$, $\varphi_3$, $\chi_2$ and $\chi_3$ obtained by appropriate interchange of indices.

Note that in the planar limit the parametrization is symmetrical between the electric charges $\gd_I$ and magnetic charges $\gc_I$.  It is possible to periodically identify $x$ and $y$, in which case the horizons are toroidal.

As for spherical horizons, various special cases with planar horizons have been given before \cite{Duff:1999gh, Duff:1999rk, Lu:2013eoa, Lu:2013ura}.  The planar solution for Einstein--Maxwell theory has been known for a long time, see e.g.\ \cite{Stephani:2003tm}, and much used in the AdS/CFT correspondence.


\subsection{Hyperbolic horizon}


From the static solution with a spherical horizon, we may perform the analytic continuations $(t, r, \gq, m, \gd_I, \gc_I) \rightarrow \im (t, r, \gq, - m, \gd_I, \gc_I)$ to obtain a solution with a hyperbolic horizon.  Conversely, for the hyperbolic solution, $(t, r, \gq, m, \gd_I, \gc_I)$ are analytic continuations of the spherical $- \im (t, r, \gq, - m, \gd_I, \gc_I)$, so it is convenient in the hyperbolic case to define $(n_0, \gz^I, \wtd{\gz}_I)$ to be the analytic continuations of the spherical $\im (n_0, \gz^I, \wtd{\gz}_I)$.  We also define $W^2, Q_I, P^I, \xi_{i 1}, \xi_{i 2}, \xi_{i 3}, \eta_{i 1}, \eta_{i 2}, \eta_{i 3}$ to be the analytic continuations of their spherical counterparts.  $W$ is defined as the positive square root of $W^2$, before or after analytic continuation.

The solution with a hyperbolic horizon has the metric
\begin{align}
\ds & = - f \, \df t^2 + \fr{\df r^2}{f} + W (\df \gq^2 + \sinh^2 \gq \, \df \phi^2) ,
\end{align}
where
\be
f = \fr{r^2}{W} \bigg( -1 - \fr{2 m}{r} + \fr{n_0^2}{r^2} + g^2 W^2 \bigg) .
\ee
The gauge fields and dual gauge fields are
\begin{align}
A^I & = \gz^I \, \df t + P^I \cosh \gq \, \df \phi , & \wtd{A}_I & = \wtd{\gz}_I \, \df t - Q_I \cosh \gq \, \df \phi .
\end{align}
The scalars are
\begin{align}
\chi_i & = \frac{f_i}{r^2 + n_0^2 + g_i} , & \expe{\gvf_i} & = \frac{r^2 + n_0^2 + g_i}{W} ,
\end{align}
where
\begin{align}
f_{i} & = 2 (n_0^2 - m r) \xi_{i 1} + 2 n_0 (r + m) \xi_{i 2} + 4 (m^2 + n_0^2) \xi_{i 3} , \nnr
g_{i} & =2 (n_0^2 - m r) \eta_{i 1} + 2 n_0 (r + m) \eta_{i 2} + 4 (m^2 + n_0^2) \eta_{i 3} .
\end{align}

Some special cases have been considered previously.  As for the spherical case, the purely electric solutions with $\gc_I = 0$ and the purely magnetic solutions with $\gd_I = 0$ were given in \cite{Duff:1999gh, Duff:1999rk} (except in our parameterization, real $\gd_I$ and $\gc_I$ give real gauge fields).  Similarly, there are the dyonic solutions \cite{Lu:2013eoa, Lu:2013ura}.  The hyperbolic version of Reissner--Nordstr\"{o}m--AdS, with $\gd_I = \gd$ and $\gc_I = \gc$, has been known for a long time, see e.g.\ \cite{Stephani:2003tm}, but its physics was studied later \cite{Aminneborg:1996iz, Mann:1996gj, Mann:1997zn, Brill:1997mf,Emparan:1999gf}.


\subsection{Thermodynamics}
\label{thermo}



\subsubsection{Mass}
\label{secmass}


In asymptotically flat spacetimes, the mass can be defined unambiguously from the Arnowitt--Deser--Misner (ADM) surface charge or from a Komar integral, and is evaluated as $M/G$ where $G$ is Newton's constant and  $M$ is written in \eqref{mass}. In asymptotically AdS spacetimes, the computation of the mass presents several subtleties (see e.g.\ \cite{Gibbons:2004ai,Chen:2005zj}). The Abbott--Deser formula \cite{Abbott:1981ff}, valid in Einstein gravity with a cosmological constant, is in general not valid when slowly falling scalar fields are present. In the example of the static solution of $\cN = 2$ gauged supergravity with four independent electric charges, it was observed that the mass is not given by the Abbott--Deser formula \cite{Chen:2005zj, Lu:2013ura}. In fact, the contributions of the scalar fields to the mass can be evaluated explicitly and uniquely (see e.g.\ \cite{Henneaux:2006hk,Barnich:2002pi,Compere:2009dp}) for any stationary solution, using either Hamiltonian \cite{Regge:1974zd,Brown:1986ed} or Lagrangian \cite{Abbott:1981ff,Iyer:1994ys,Barnich:2001jy,Barnich:2007bf} canonical methods. Holographic methods \cite{Papadimitriou:2005ii} could also be used. Here we simply implement the canonical methods on the theory at hand and uniquely compute the mass. The explicit formulae for the conserved charges of a Lagrangian of the form \eqref{generalaction} can be found in \cite{Compere:2009dp}.

From now on, we set the NUT charge to zero and consider spherical horizons. In order to perform the asymptotic analysis, it is convenient to first expand $W^2(r)$ and the scalar fields $(\varphi_i,\chi_i)$, as
\begin{align}
W^2(r) & = r^4 + w_3 m r^3 + w_2 m^2 r^2 + w_1 m^3 r + w_0 m^4,\nn \\
\varphi_i & = \frac{\Sigma_i}{r}+\frac{\Sigma_i^{(2)}}{r^2}+O(r^{-3}),\nnr
\chi_i & = \frac{\Xi_i}{r}+\frac{\Xi^{(2)}_i}{r^2}+O(r^{-3}).
\label{subsc}
\end{align}
The various coefficients can be obtained from the definitions \eqref{defW2} and \eqref{defsc}. It is only important to notice the relationships
\begin{align}
w_3 & = 4 \bigg( \frac{M}{m} - 1\bigg) , \\
w_2 & = \frac{3}{8}w_3^2-\frac{1}{2m^2}\sum_i (\Sigma_i^2 + \Xi_i^2) ,\label{vald}\\
w_1 & = \frac{1}{16}w_3^3 - \frac{5}{12m^2} w_3 \sum_i (\Sigma_i^2 + \Xi_i^2) \nnr
& \quad -\frac{1}{3m^3}\sum_i (\Sigma_i \Xi_i^2 + 2\Sigma^{(2)}_i \Sigma_i + 2 \Xi^{(2)}_i \Xi_i) ,
\end{align}
which can be checked explicitly.

The contribution to an infinitesimal variation of the mass coming from the Einstein term in the action can be obtained with the result
\be
G \, \delta M_{\textrm{Einstein}} = \frac{g^2}{32} \delta \big[ (8w_2-3w_{3}^2)m^2 \big] r +O(r^0).
\label{dM1}
\ee
The linear divergence in $r$ is a clear sign that matter fields are contributing to the mass. One can easily check that the gauge fields are not contributing to the mass using the explicit formulae for the conserved charges \cite{Compere:2009dp}. Since the magnetic monopole fields do not enter the expression for the mass at radial infinity, there is no subtlety associated with Dirac strings or singularities of the gauge fields. The scalar field contribution reads
\begin{align}
G \, \delta M_{\textrm{scalars}} & = - \frac{1}{16\pi }\sum_i \int \! \df \theta \, \df \phi \, \sqrt{-g} g^{r r} ( \delta \varphi_i \p_r \varphi_i \nnr
& \quad + \expe{2\varphi_i} \delta \chi_i \p_r \chi_i ) .
\end{align}
For asymptotically flat black holes, $\delta M_{\textrm{scalars}} =O(r^{-1})$, and the scalars do not contribute to the mass. For asymptotically AdS black holes, there is an additional quartic branch in $g^{rr}$ and we obtain a linearly divergent term as well as a finite term:
\be
G \, \delta M_{\textrm{scalars}} = \frac{g^2 r}{8} \sum_i \delta (\Sigma_i^2+\Xi_i^2) +O(r^0) .
\label{dM2}
\ee
Summing up the two contributions \eqref{dM1} and \eqref{dM2} and using \eqref{vald}, the linearly divergent terms cancel. The variation of the mass then takes the finite form,
\begin{align}
G \, \delta \mathcal M &= \delta M +\frac{g^2}{12} \sum_i [ - (\Xi_i^2 +\Sigma_i^2) \delta (M- m) \nnr
& \quad + \tfrac{1}{2}(M-m) \delta (\Xi_i^2 + \Sigma_i^2) + 2 \Xi_i^{(2)} \delta \Xi_i - \Xi_i \delta \Xi_i^{(2)} \nonumber \\
& \quad + 2\Sigma_i^{(2)} \delta \Sigma_i - \Sigma_i \delta \Sigma_i^{(2)} +\Sigma_i \delta (\Xi_i^2) - 2 \Xi_i^2 \delta \Sigma_i ] .
\end{align}
The variations proportional to $\delta m$ exactly cancel in the last term. Therefore, if the mass is integrable, its value will be $\mathcal M = M/G$. The integrability condition then amounts to
\be
\mathcal I = 0 ,
\ee
where
\begin{align}
\mathcal I & \equiv g^2 \sum_i [ \delta_2 (\Sigma^{(2)}_i - \Xi_i^2) \wedge \delta_1 \Sigma_i + \delta_2 \Xi^{(2)}_i \wedge \delta_1 \Xi_i \nnr
& \quad -\tfrac{1}{2} \delta_2 (\Xi_i^2 + \Sigma_i^2) \delta_1 (M-m) - (1\leftrightarrow 2) ] .\label{II}
\end{align}
Let us now discuss the physical content of the integrability condition. It turns out that for generic values of the electric and magnetic charge parameters, the integrability condition is not obeyed, and therefore, the mass does not exist. The non-integrability of the mass comes more precisely from the presence of non-trivial scalar field profiles when the gauge fields with both electric and magnetic charges are turned on. This non-integrability is rooted in the non-existence of a phase space with such generic variations. Indeed, one can evaluate the symplectic flux \eqref{sympl} on constant $r$ slices when $r\rightarrow \infty$ using \eqref{vald}. The result is
\be
\bm\omega (\delta_1 \phi,\delta_2\phi) |_{r \;\text{fixed}}= \mathcal{I} \sin \theta \, \df t\wedge \df \theta \wedge \df \phi +O(r^{-1}),
\ee
where $\mathcal I$ is given in \eqref{II}. One fundamental requirement of consistent boundary conditions is that the symplectic form has to be conserved, i.e.\ the symplectic flux at infinity should be zero. Therefore, the mass does not have to exist, since there is no classical phase space that contains generic independently varying electric and magnetic charges. In the case studied in \cite{Lu:2013ura}, where one $\textrm{U}(1)$ gauge field is turned on, one can also check that the mass is not generically defined, which follows from the non-existence of a conserved symplectic structure.

In the cases where we expect a phase space to be defined, such as only electric charges (all $P^I = 0$) or only magnetic charges (all $Q_I = 0$), where a dual CFT description should be available \cite{Witten:2003ya}, the mass should also be defined, and indeed it is. One can check that $\mathcal I=0$ for these cases, and the mass is then $\cM = M/G$.

Quite surprisingly, there are also cases with independent electric and magnetic charges where the symplectic flux vanishes ($\mathcal I=0$), and the mass is defined. These boundary conditions will violate boundary Lorentz invariance and therefore will be outside of the standard AdS/CFT description. One such example is when the gauge fields are pairwise equal, e.g.\ $Q_1=Q_4$, $Q_2=Q_3$, $P^1=P^4$, $P^2=P^3$ (which is equivalent to $\delta_1 = \delta_4$, $\delta_2 =\delta_3$, $\gamma_1 = \gamma_4$, $\gamma_2=\gamma_3$). This includes as a subcase the dyonic Reissner-Nordtr\"om--AdS black hole.

Quite intriguingly, at least three additional distinct cases arise where the integrability conditions $\mathcal I=0$ are obeyed. The first case is when all four electric and magnetic charges are set equal or opposite to each other $P^I = \pm Q_I$,  $I=1,2,3,4$, with an even number of minus signs. The second case arises with two vanishing sets of charges, let say $Q_1=Q_2=P^1=P^2=0$, and with the two remaining sets of charges obeying $Q_3 = \pm P^3$, $Q_4 = \pm P^4$ for any choice of signs. The third case arises when all but one set of charges vanishes, let say $Q_1$ and $P^1$, with $Q_1 = \pm P^1$. The last case was discussed in \cite{Lu:2013ura}. For free gauge fields in AdS$_4$, one has an $\SL(2,\mathbb Z)$ family of Lorentz-invariant boundary conditions \cite{Witten:2003ya}. Here, we find a smaller set of mixed boundary conditions, which are very restricted by the interactions with the scalar fields. Other examples might exist, since we were not able to perform a complete classification of the solutions to $\mathcal I=0$.

\subsubsection{Other conserved charges}
\label{consch}

Let us obtain the electromagnetic charges and the angular momentum of the solution. Using the canonical expressions for the charges \cite{Compere:2009dp} for the Lagrangian \eqref{generalaction}, one obtains the electromagnetic charges in geometrical units, $\overline{Q}_I = \tf{1}{4G} Q_I$ and $\overline{P}^I = \tf{1}{4G} P^I$. Since electric and magnetic charges are present, the definition of angular momentum requires more care.

The angular momentum is usually associated with the large gauge transformation $(\xi,\Lambda^I) = (- \pd / \pd \phi,0)$, where $\xi$ is the generator of a diffeomorphism, and $\Lambda^I$ are generators of $\textrm{U}(1)$ gauge transformations. However, in the presence of magnetic charges, the definition has to be modified. Remember that in the presence of magnetic charges, the gauge field $A^I$ might be defined in the north and south patches as
\begin{align}
A_{\textrm{north}}^I & = P^I (\cos\theta -1) \, \df \phi +O(r^{-1}), \nnr
A_{\textrm{south}}^I & = P^I (\cos\theta +1) \, \df\phi + O(r^{-1}).
\end{align}
One requirement from the applicability of the canonical formalism at $r \rightarrow \infty$ is that $\xi^\mu A^I_\mu + \Lambda^I$ has to be continuous across the equator at $r \rightarrow \infty$ \cite{Copsey:2005se}. We define the magnetic charges of the gauge field $\mathbf A^I$ as defined in \eqref{generalaction} by $\mathbf{P}^I$, $I=1,2,3,4$, namely $\mathbf{P}^1 = P^1$, $\mathbf{P}^2 = -Q_2$, $\mathbf{P}^3 = -Q_3$ and $\mathbf{P}^4 = P^4$. The angular momentum can then be defined as associated with  $(-\pd / \pd \phi,\Lambda^I)$ where $\Lambda^I_{\textrm{north}}=-\mathbf{P}^I$, $\Lambda^I_{\textrm{south}}= \mathbf{P}^I$. One can then check that the angular momentum is zero. Indeed, one has explicitly
\be
\delta J = \int \! \frac{(\df ^2 x)_{\mu\nu}}{16\pi G} \, \left[  \delta T_I^{\mu \nu} (\mathbf A^I_\phi +\Lambda^I ) + T_I^{\mu\nu} \delta \mathbf A_\phi^I \right] ,
\ee
where $T_I^{\mu\nu} = k_{IJ}\mathbf F^{J \mu \nu}-h_{IJ}\eps^{\mu\nu\lambda\sigma}\mathbf F_{\lambda \sigma}^J$. Evaluating on the sphere at infinity $S^2_\infty$ and using
\begin{align}
k_{IJ} & = \delta_{IJ} + O(r^{-1}), & h_{IJ} & = O(r^{-1}),
\end{align}
one then has
\be
\delta J = \fr{1}{16 \pi G} \int_{S^2_\infty} \! \df \theta \, \df \phi \, r^2  \mathbf F_I^{tr} \delta \mathbf A_\phi^I = 0 ,
\ee
after a non-trivial cancellation between the north and south patches. Physically, since the magnetic monopole sits on the electric monopole, no net angular momentum is produced.


\subsubsection{First law}
\label{fl1}


Black hole solutions have horizons at $r = r_\pm$, which are the roots of the polynomial $R_g(r)$ defined as
\be
R_g(r) = R(r) + g^2 W^2(r) ,
\ee
where $R$ and $W$ are given in \eqref{defW2} and \eqref{defR}.  From the geometry, we obtain the temperature $T$ and entropy $S$
\begin{align}
T & = \fr{R_{g}' (r_+)}{4 \pi W(r_+)} , & S & = \frac{\pi}{G} W(r_+) ,
\end{align}
where prime denotes radial derivative.  The electric and magnetic potentials on the horizon are respectively
\begin{align}
\Phi^I & =  \gz^I (r_+) , & \Psi_I & = \wtd{\gz}_I (r_+).
\end{align}
In the asymptotically flat case ($g=0$), these expressions can be simplified after using the property $W(r_+)=L(r_+)$ as
\begin{align}
\Phi^I |_{\textrm{flat}} & = -\frac{P^I n_0+L^I(r_+)}{L(r_+)}, & \Psi_I|_{\textrm{flat}} & = \frac{Q_I n_0+\widetilde L_I(r_+)}{L(r_+)},
\end{align}
but there is no obvious simplification in the asymptotically AdS case.

For each boundary condition where we explicitly checked that the mass can be defined, we find that the thermodynamic quantities satisfy the first law of thermodynamics
\ben
\gd \mathcal M = T \, \gd S + \sum_{I = 1}^4 ( \Phi^I \, \gd \overline{Q}_I + \Psi_I \, \gd \overline{P}^I) ,
\een
where $\overline{Q}_I = \tf{1}{4G} Q_I$ and $\overline{P}^I = \tf{1}{4G} P^I$ are the electromagnetic charges in geometrical units obtained earlier.

We finally note that the thermodynamics may be studied similarly for the solutions with planar or hyperbolic horizons.  The thermodynamic quantities for the planar solutions can easily be read off from the solution.  The thermodynamic quantities for hyperbolic horizons can be obtained from those of the solution with a spherical horizon by analytic continuation.


\section{$\textrm{U}(1)^2$ dyonic rotating black holes}
\label{BH2}


Let us now present rotating solutions to the Lagrangian with the four gauge fields set pairwise equal, \eqref{LU2} or equivalently \eqref{SO4}. The general stationary rotating black hole of $\cN = 2$, $\textrm{U}(1)^2$ ungauged supergravity was found in \cite{LozanoTellechea:1999my}, and is asymptotically flat or more generally asymptotically Taub--NUT. A generalization of this black hole to $\cN = 2$, $\textrm{U}(1)^4$ ungauged supergravity was presented in a simplified form in \cite{Chow:2013tia}. Using the $(r, u)$ symmetric notation \cite{Chong:2004na}, the metric can be written as
\begin{align}
\ds & = - \fr{\widehat R}{\widehat W} \bigg( \df \widehat  t - \fr{\widehat a^2 - \widehat u_1 \widehat u_2}{\widehat a} \, \df \widehat \gf \bigg)^2 + \fr{\widehat W}{\widehat{R}} \, \df \widehat r^2 \nnr
& \quad + \fr{\widehat U}{\widehat W} \bigg( \df \widehat t - \fr{\widehat r_1 \widehat r_2 + \widehat a^2}{\widehat a} \, \df \widehat \gf \bigg)^2 + \fr{\widehat{W}}{\widehat{U}} \, \df \widehat u^2 ,
\end{align}
where $\widehat W=\widehat r_1 \widehat r_2 + \widehat u_1 \widehat u_2$; $\widehat R(\widehat r)$ and $\widehat U(\widehat u)$ are quadratic polynomials of their arguments; and $\widehat r_a,\widehat u_a$ for $a=1,2$ are linear functions of $\widehat r, \widehat u$, respectively. One can think of $\widehat u$ as a function of $\cos\theta$, where $\theta$ is the polar angle. Following \cite{Chong:2004na}, it turns out that one can obtain a solution to $\textrm{U}(1)^2$ gauged supergravity upon replacing the functions as
\begin{align}
\widehat R & \rightarrow \widehat R+g^2 \widehat r_1 \widehat r_2 (\widehat r_1 \widehat r_2 +\widehat a^2), \\
\widehat U &\rightarrow \widehat U+g^2 \widehat u_1 \widehat u_2 (\widehat u_1 \widehat u_2 -\widehat a^2),
\end{align}
with everything else untouched, including the matter fields. In order to set the metric in an asymptotically AdS coordinate system, an analysis \emph{\`a la} Griffiths--Podolsk\'y \cite{Griffiths:2005qp} is necessary, which we will perform in Section \ref{AAdS}.  Since the interest of these solutions is the inclusion of angular momentum, we will only discuss the black hole solutions with spherical horizons.


\subsection{General solution}



\subsubsection{Metric}


We use hatted coordinates to emphasize that the solution is not in an asymptotically AdS coordinate system.  The coordinates are rotating at infinity and $\widehat{\phi}$ is not canonically normalized.  However, this coordinate system is convenient for expressing the solution in a simple form. Using the notations of \cite{Chong:2004na}, the metric is
\begin{align}
\ds & = - \fr{\widehat R_g}{\widehat W} \bigg( \df \widehat  t - \fr{\widehat a^2 - \widehat u_1 \widehat u_2}{\widehat a} \, \df \widehat \gf \bigg)^2 + \fr{\widehat W}{\widehat R_g} \, \df \widehat r^2 \nnr
& \quad + \fr{\widehat U_g}{\widehat W} \bigg( \df \widehat t - \fr{\widehat r_1 \widehat r_2 + \widehat a^2}{\widehat a} \, \df \widehat \gf \bigg)^2 + \fr{\widehat{W}}{\widehat U_g} \, \df \widehat{u}^2 ,\label{mads3}
\end{align}
where
\begin{align}
\widehat R_g(\widehat r) & = \widehat  r^2 - 2 \widehat m \widehat r + \widehat a^2  + g^2 \widehat r_1 \widehat r_2 (
\widehat r_1 \widehat r_2 + \widehat a^2) , \nnr
\widehat U_g(\widehat u) & = -\widehat u^2 + 2 \widehat n \widehat u +\widehat a^2 + g^2 \widehat u_1 \widehat u_2 (\widehat u_1 \widehat u_2 - \widehat a^2) , \nnr
\widehat W(\widehat r,\widehat u) & = \widehat r_1 \widehat r_2 + \widehat u_1 \widehat u_2 . \label{hRg}
\end{align}
The variables $\widehat r_a$, $\widehat u_a$ are defined as
\begin{align}
\widehat r_a & = \widehat r + \gD \widehat r_a , & \widehat u_a & = \widehat u + \gD \widehat u_a ,
\end{align}
where
\begin{align}
\gD \widehat r_1 & = \widehat m [\cosh (2 \delta_1) \cosh (2 \gamma_2) - 1] \nnr
& \quad + \widehat n \sinh (2 \delta_1) \sinh (2 \gamma_1) , \nnr
\gD \widehat r_2 & = \widehat m [\cosh (2 \delta_2) \cosh (2 \gamma_1) - 1] \nnr
& \quad + \widehat n \sinh (2 \delta_2) \sinh (2 \gamma_2) , \nnr
\gD \widehat u_1 & = \widehat n [\cosh (2 \delta_1) \cosh (2 \gamma_2) - 1] \nnr
& \quad - \widehat m \sinh (2 \delta_1) \sinh (2 \gamma_1) ,\nnr
\gD \widehat u_2 & = \widehat n [\cosh (2 \delta_2) \cosh (2 \gamma_1) - 1] \nnr
& \quad - \widehat m \sinh (2 \delta_2) \sinh (2 \gamma_2) .
\end{align}
It is convenient to define the linear combinations of $\Delta \widehat r_a,\Delta \widehat u_a$, as
\begin{align}
\Sigma_{\Delta r} & = \tfrac{1}{2} (\Delta \widehat r_1 + \Delta \widehat r_2) , & \Delta_{\Delta r} & = \tfrac{1}{2} (\Delta \widehat r_2 - \Delta \widehat r_1) , \nnr
\Sigma_{\Delta u} & = \tfrac{1}{2} (\Delta \widehat u_1 + \Delta \widehat u_2) , & \Delta_{\Delta u} & = \tfrac{1}{2} (\Delta \widehat u_2 - \Delta \widehat u_1) . \label{SigmaDelta}
\end{align}
We define
\begin{align}
M & = \widehat m + \Sigma_{\Delta r} , & N & = \widehat n + \Sigma_{\Delta u} ,
\end{align}
which are the physical mass and NUT charge in the asymptotically flat case when no gauging is present ($g=0$). It turns out that the total NUT charge in asymptotically AdS spacetime vanishes when $N=0$ as we will explicitly show in Section \ref{AAdS}. The NUT charge can therefore be cancelled upon setting $\widehat n=\widehat n_0$, where
\be
\widehat n_0 = \fr{\sinh (2 \gamma_1) \sinh (2 \delta_1) + \sinh (2 \gamma_2) \sinh (2 \delta_2)}{\cosh (2 \gamma_1) \cosh (2 \delta_2) + \cosh (2 \gamma_2) \cosh (2 \delta_1)} \widehat m.\label{noNUT1}
\ee
We fix the orientation as $\varepsilon_{t r \phi u} =1$.


\subsubsection{Matter}


The electromagnetic charges are
\begin{align}
Q_1 & = \frac{\p M}{\p \delta_1} = \frac{1}{2}\frac{\p \widehat r_1}{\p \delta_1}, &Q_2 & = \frac{\p M}{\p \delta_2} = \frac{1}{2}\frac{\p \widehat r_2}{\p \delta_2}, \nnr
P^1 & = - \frac{\p N}{\p \delta_1} = - \frac{1}{2}\frac{\p \widehat u_1}{\p \delta_1}, & P^2 & = - \frac{\p N}{\p \delta_2} = - \frac{1}{2}\frac{\p \widehat u_2}{\p \delta_2} .
\end{align}
The gauge fields and dual gauge fields are
\begin{align}
A^{1} & = {\zeta}^1 (\df\widehat  t -  \widehat a \, \df\widehat \phi) + \frac{\widehat r_2 \widehat u_2  \wtd\zeta_1}{\widehat a}\df\widehat \phi,  \nnr
A^{2} & = {\zeta}^2 (\df \widehat  t -\widehat  a \,\df \widehat \phi) + \frac{\widehat r_1 \widehat u_1  \wtd\zeta_2}{\widehat a}\df\widehat \phi , \nnr
 \wtd{A}_1 & = \wtd\gz_1 (\df \widehat t - \widehat a \, \df \widehat \gf) - \fr{\widehat r_1 \widehat u_1 {\zeta}^1}{\widehat  a}  \, \df \widehat \gf , \nnr
 \wtd{A}_2 & = \wtd\gz_2 (\df \widehat  t -  \widehat a \, \df \widehat \gf) - \fr{\widehat r_2 \widehat u_2 {\zeta}^2}{\widehat  a} \, \df \widehat \gf ,
\end{align}
where the 3-dimensional electromagnetic scalars are
\begin{align}
{\gz}^1 & = \fr{1}{2 \widehat W} \fr{\pd \widehat W}{\pd \delta_1} = \fr{Q_1 \widehat r_2 - P^1 \widehat u_2}{\widehat W}, & \wtd\gz_1 & = \fr{Q_1 \widehat u_1 + P^1 \widehat r_1}{\widehat W}, \nnr
{\gz}^2 & = \fr{1}{2 \widehat W} \fr{\pd \widehat W}{\pd \delta_2} = \fr{Q_2\widehat r_1 - P^2 \widehat u_1 }{\widehat W} , & \wtd\gz_2 & = \fr{Q_2 \widehat u_2 + P^2 \widehat r_2}{\widehat W} .
\end{align}
Here, partial differentiation is done for generic $\widehat n$ and the result is then evaluated on $\widehat n=\widehat n_0$.  The scalar fields are given by
\begin{align}
\chi & = \frac{\widehat r_2 \widehat u_1 - \widehat r_1 \widehat u_2}{\widehat r_2^2 + \widehat u_2^2} , & \expe{\varphi} & = \frac{\widehat r_2^2+\widehat u_2^2}{\widehat W} . \label{scalarspairwise}
\end{align}
It is quite remarkable that the gauge fields have such simple expressions in terms of the scalars $\zeta^I$ and $\wtd \zeta_I$, and that the scalars themselves are simple.  We checked that the metric \eqref{mads3}, accompanied with the matter fields, is a solution to the field equations of the Lagrangian \eqref{LU2} using Mathematica \cite{Mathematica}.


\subsection{Asymptotically AdS coordinates}
\label{AAdS}


We have presented a local form of the metric and the matter fields. The identification of the total NUT charge and the global identifications of coordinates necessary in order to have a regular solution can be obtained by finding a suitable asymptotically AdS coordinate system in the asymptotic region. We follow the method of Griffiths and Podolsk\'y \cite{Griffiths:2005qp}, which consists of setting the metric in a suitable generalization of the Pleba\'{n}ski--Demia\'{n}ski form for which the analysis can be done most easily.

Starting from the coordinates $(\widehat r,\widehat u,\widehat \phi,\widehat t)$, we define the new coordinates $(r,p,\overline \phi,t)$ as
\begin{align}
\widehat r & = \beta r - \Sigma_{\Delta r} , \\
\widehat \phi & = \frac{\widehat a}{\beta^3 a}\overline \phi,\\
\widehat u & = \beta (N_g + a p) -\Sigma_{\Delta u} ,\\
\widehat t & = \frac{t}{\beta} + \frac{\widehat a^2 + \Delta_{\Delta u}^2 -(N_g+a)^2 \beta^2 }{a \beta^3} \overline \phi
\end{align}
where $a,\;\beta,\; N_g$ are three constants that we will fix shortly in terms of the parameters of the solution. The metric then reads as
\begin{align}
\ds & = -\frac{R_g}{W} \big( \df t - [2N_g (1- p) + a (1-  p^2) ] \, \df \overline \phi \big) ^2 \nnr
& \quad + \frac{P_g}{W} \big( a \, \df t - [r^2 - v^2 + (N_g + a)^2] \, \df \overline \phi \big)^2 \nn\\
& \quad + W  \bigg( \fr{\df r^2}{R_g} + \fr{\df p^2}{  P_g} \bigg) ,
\end{align}
with
\begin{align}
W & = r^2 + (N_g + a  p)^2 - v^2 , \nnr
R_g & = k + e^2 -2 m r +(\epsilon -2 g^2 v^2)r^2+g^2 r^4, \nnr
P_g & = a_0 +a_1  p+a_2  p^2+a_3  p^3 +a_4  p^4,
\end{align}
where
\begin{align}
a_0 &= a^{-2}(k -N_g^2 \eps +2 n N_g +g^2 N_g^4),\nnr
a_1 &= 2a^{-1}(n-N_g \eps +2g^2 N_g^3),\nnr
a_2 &= 6 g^2 N_g^2 -\eps,\nnr
a_3 &= 4 a g^2 N_g,\nnr
a_4 &= g^2 a^2,
\end{align}
and one can express the new parameters $(\epsilon,k,e,m,n,v)$ in terms of the old ones as
\begin{align}
m &= \beta^{-3}(\widehat m+\Sigma_{\Delta r}),\label{valm}\\
n & = \beta^{-3} (\widehat n+\Sigma_{\Delta u} ) , \label{valn}\\
v^2 &= \beta^{-2} (\Delta_{\Delta r}^2 +\Delta_{\Delta u}^2  ),\label{vv}\\
\epsilon & = \beta^{-2} [1 + g^2 (\widehat  a^2+2\Delta^2_{\Delta u}  ) ] , \label{valeps}\\
k & = \beta^{-4} [\widehat a^2 - 2 \widehat{n} \Sigma_{\Delta u} - \Sigma_{\Delta u}^2 + g^2 \Delta_{\Delta u}^2 (\widehat a^2+\Delta_{\Delta u}^2 )] , \label{valk}\\
e^2 +k & = \beta^{-4} [\widehat a^2+2 \widehat m \Sigma_{\Delta r} + \Sigma_{\Delta r}^2 - g^2\Delta_{\Delta r} ^2 (\widehat a^2-\Delta_{\Delta r}^2)] .\label{vale}
\end{align}
We have set the metric in a form where the treatment of \cite{Griffiths:2005qp} is applicable. Note that our metric is slightly more general than the non-accelerating metric studied in \cite{Griffiths:2005qp} (obtained by setting their $\alpha$ to zero), since here $v$ is generically non-zero. In generalizing the Griffiths--Podolsk\'y form, we insisted on keeping the same dependence for $p$-dependent functions, since this dependence is used to discuss the range of the angular coordinates of the solutions. Instead, the radial dependence is changed by $O(r^{-2})$ terms that do not affect the analysis.

When $P_g$ has two real roots (which is the case of interest here), one can choose to put these real roots at $1$ and $-1$,
\be
P_g = (1- p^2)(a_0 -a_3 p-a_4 p^2),
\ee
which implies that the above coefficients obey $a_1+a_3 =0 = a_0+a_2+a_4$. These two conditions provide two linear equations that specify $\eps$ and $n$ in terms of $a$ and $N_g$ as
\begin{align}
\eps & = \frac{k}{a^2-N_g^2}+g^2 (a^2+3N_g^2),\label{eqepsn0} \\
n & = \frac{k N_g}{a^2-N_g^2}+g^2 N_g(N_g^2-a^2).\label{eqepsn}
\end{align}
The parameter $N_g$ is then recognized as the total NUT charge since the metric admits a NUT singularity proportional to $N_g$ at the pole $p = -1$. The condition that the NUT charge $N_g$ is zero is equivalent to $n=0$, which from \eq{valn} translates into
\be
\widehat{n} + \Sigma_{\Delta u} = 0 ,
\ee
which is equivalent to \eqref{noNUT1}. This justifies therefore our previous claim.

Let us assume $a_0 > 0$. This assumption can be verified for all Pleba\'{n}ski--Demia\'{n}ski black holes and for our general black holes when $N_g=0$.  We can then set $a_0= 1$ using the remaining scaling symmetry. The equation $a_0 = 1$ can be solved for $k$ as
\be
k = (a^2 - N_g^2)(1+3g^2 N_g^2). \label{solk}
\ee
Hence, equations \eqref{eqepsn0} and \eqref{eqepsn} become
\begin{align}
\eps & = 1 +g^2 (a^2 + 6 N_g^2), \label{soleps} \\
n & = N_g [1 -g^2(a^2-4N_g^2)] .
\end{align}
After setting $p =\cos\theta$ and $\overline\phi = \widetilde{\phi}\, \Xi^{-1}$, where
\be
\Xi = 1-a^2 g^2-4 a N_g g^2,
\ee
the metric in $(t,r,\theta,\widetilde{\phi})$ coordinates becomes
\begin{align}
\ds & = -\frac{R_g}{W}\left( \df t - \frac{a \sin^2 \theta + 4 N_g \sin^2(\theta/2)}{\Xi}\, \df \widetilde{\phi} \right)^2 \nnr
& \quad + \frac{\Theta_g \sin^2\theta}{W }\left(a \, \df t - \frac{L}{\Xi} \, \df \widetilde{\phi} \right)^2 + W  \bigg( \fr{\df r^2}{R_g} + \fr{\df \theta^2}{\Theta_g} \bigg) ,\label{AAdSm}
\end{align}
with
\begin{align}
R_g(r) & = r^2 - 2 m r + a^2 + e^2 - N_g^2 + g^2 [ r^4 \nnr
& \quad + (a^2 + 6 N_g^2-2v^2 )r^2+ 3N_g^2(a^2-N_g^2) ],\label{Rg}\\
\Theta_g(\theta) & = 1 - a^2 g^2 \cos^2\theta - 4 a g^2 N_g \cos\theta , \\
W(r,\theta) & = r^2 + (N_g+a\cos\theta )^2 - v^2, \label{defW} \\
L(r) & = r^2 +(N_g+a)^2 - v^2,\label{defL}
\end{align}
which is a straightforward generalization of the Kerr--Newman--Taub--NUT--AdS solution. The solution is regular at the north and south poles $\theta = 0,\pi$ upon identifying $\widetilde{\phi} \sim \widetilde{\phi} + 2\pi$ with $0 \leq \theta \leq \pi$.

When the NUT charge is zero, the $(t, \widetilde{\phi})$ coordinate frame is rotating at infinity, but provides a concise way of stating the solution.  To obtain a coordinate frame that is static at infinity, we should use
\be
\phi = \widetilde{\phi} + a g^2 t ,
\label{Omegainf}
\ee
which also has period $2 \pi$.  Furthermore, manifestly asymptotically AdS coordinates $(t,r_*,\theta_*,\phi)$ can be obtained from $(t,r,\theta,\phi)$ coordinates as
\begin{align}
r & = r_* \sqrt{1 - a^2 g^2 \sin^2\theta_*} \bigg( 1 \nnr
& \quad + \frac{v^2-a^2\sin^2\theta_* [1-g^2 (v^2-a^2)]}{2(1 - a^2 g^2 \sin^2\theta_*)^2 r_*^2}+O(r_*^{-4}) \bigg) , \nnr
\sin\theta & = \frac{\sr{\Xi}}{\sr{1-a^2 g^2 \sin^2\theta_*}}\sin\theta_* \bigg( 1 \nnr
& \quad - \frac{a^2 \cos\theta_*}{2(1-a^2 g^2 \sin^2\theta_*)^2 r_*^2}+O(r_*^{-4})\bigg) ,
\end{align}
The metric reads as
\begin{align}
\ds & = -(1 + g^2 r_*^2) \df t^2 + \frac{\df r_*^2}{1 + g^2 r_*^2} + r_*^2 (\df \theta_*^2 +\sin^2\theta_* \, \df \phi^2) \nnr
& \quad + h_{\mu\nu} \, \df x^\mu \, \df x^\nu \label{adsr}
\end{align}
with
\begin{align}
h_{t t} & = O(r_*^{-1}), \quad h_{t \phi} = O(r_*^{-1}),\quad h_{\phi \phi} = O(r_*^{-1}),\nnr
h_{t r_*} & = h_{t_* \theta_*} = h_{\phi r_*} = h_{\phi \theta_*} =0,\nnr
h_{r_* r_*} & = -\frac{v^2}{g^2(1-g^2 a^2 \sin^2\theta_*)r_*^4}+O(r_*^{-5}),\nnr
h_{r_* \theta_*} & = O(r_*^{-3}),\qquad  h_{\theta_* \theta_*} = O(r_*^{-2}).
\end{align}
and the coordinates are identified as $0 \leq \theta_* \leq \pi$. This completes our program of finding a global coordinate system for the solution in the asymptotic region.

In the presence of scalar fields, asymptotically AdS boundary conditions are generically modified while the asymptotic symmetry group is generically unchanged, see e.g.\ \cite{Hertog:2004dr,Henneaux:2006hk}. Here, when $v \neq 0$, the metric \eqref{adsr} does not obey the Henneaux--Teitelboim boundary conditions \cite{Henneaux:1985tv}. When only electric charges are allowed, we expect that the asymptotic symmetry group is still the $\SO(3,2)$ group. When both electric and magnetic charges are allowed to be varied, boundary conditions for the gauge fields violate Lorentz invariance as we discussed in the introduction. We then expect that the asymptotic symmetry group only consists of the Galilean group of rotations and translations.

In order to express the solution in terms of the original parameters $(\widehat m, \widehat n,\widehat a)$, we can compare the solutions for $(\eps,k)$ in \eqref{solk} and \eqref{soleps} with \eqref{valk} and \eqref{valeps}. Comparing $\epsilon$ shows that $\beta$ is given by
\be
\beta^2 = \frac{1+g^2(\widehat a^2+2 \Delta_{\Delta u}^2)}{1+g^2 (a^2+6N_g^2)} . \label{solbeta}
\ee
Comparing $k / \epsilon^2$ shows that $\widehat a$ and $a$ are related by a quadratic equation for $\widehat{a}^2$ in terms of $a^2$ and vice versa.  When $N_g=0$, this relation reduces to
\be
\frac{\widehat a^2 - \Sigma_{\Delta u}^2+g^2 \Delta_{\Delta u}^2 (\widehat a^2 + \Delta_{\Delta u}^2)}{[1+g^2(\widehat a^2 +2\Delta_{\Delta u}^2)]^2} = \frac{a^2}{(1+a^2 g^2)^2}.\label{solhata}
\ee
If furthermore $\Delta_{\Delta u} = 0 = \Sigma_{\Delta u}$, then the two solutions are $\widehat a^2 = a^2$ and $\widehat a^2 = a^{-2} g^{-4}$. We select the branch that is smooth in the limit $g \rightarrow 0$, which reduces in pure gravity to  $\widehat a = a$. This fixes uniquely the relationship between $\widehat a$ and $a$. The parameters $(m,n)$ are then related to $(\widehat m,\widehat n)$ by \eqref{valm} and \eq{valn} where $\beta$ is given by \eqref{solbeta}.

To summarize, the solution depends on the gauge-coupling constant $g$ and 7 other parameters: the mass parameter $m$, the NUT charge $N_g$, the rotation parameter $a$, and four electromagnetic charge parameters $\Sigma_{\Delta r}, \Delta_{\Delta r}, \Sigma_{\Delta u}, \Delta_{\Delta u}$.  From the original solution expressed with hatted parameters $(\widehat{m}, \widehat{n}, \widehat{a})$, and charge parameters $(\gd_1, \gd_2, \gc_1, \gc_2)$, it is trivial to compute $\Sigma_{\Delta r}, \Delta_{\Delta r}, \Sigma_{\Delta u}, \Delta_{\Delta u}$ from \eq{SigmaDelta}.  $a$ is then computed from \eq{solhata}, or its generalization to non-zero NUT charge.  $\beta$ is then computed from \eq{solbeta}, with which $m$ and $n$ are computed using \eq{valm} and \eq{valn}.  Finally, $v$ and $e$ can be computed using \eq{vv}, \eq{valk} and \eq{vale}.

The matter fields can then be expressed in the coordinates $(t,r,\theta,\widetilde\phi)$. In the asymptotically AdS case, setting $N_g = 0$, we notice that $\widehat W = \beta^2 W$ and
\begin{align}
\widehat r_1 & = \beta r -\Delta_{\Delta r}, & \widehat r_2 & = \beta r +\Delta_{\Delta r}, \nnr
\widehat u_1 & = \beta a \cos\theta -\Delta_{\Delta u}, & \widehat u_2 & = \beta a \cos\theta +\Delta_{\Delta u} .
\end{align}
We can immediately express the scalar fields \eq{scalarspairwise} in the new coordinates.  The gauge fields are
\begin{align}
A^{1} & = \frac{Q_1}{\beta^2 W}(r+\beta^{-1}\Delta_{\Delta r}) \bigg( \df t -\frac{a\sin^2\theta}{\Xi} \, \df \widetilde{\phi} \bigg) \nnr
& \quad - \frac{P^1}{\beta^2 W}(a\cos\theta+\beta^{-1}\Delta_{\Delta u}) \bigg(\df t - \frac{L}{\Xi a} \, \df \widetilde{\phi} \bigg) ,\nnr
A^{2} & = \frac{Q_2}{\beta^2 W}(r-\beta^{-1}\Delta_{\Delta r}) \bigg( \df t -\frac{a\sin^2\theta}{\Xi} \, \df \widetilde{\phi} \bigg) \nnr
& \quad - \frac{P^2}{\beta^2 W}(a\cos\theta - \beta^{-1}\Delta_{\Delta u}) \bigg(\df t - \frac{L}{\Xi a} \, \df \widetilde{\phi} \bigg) ,
\end{align}
and the dual gauge fields are
\begin{align}
\widetilde{A}_{1} & = \frac{P^1}{\beta^2 W}( r - \beta^{-1}\Delta_{\Delta r}) \bigg( \df t -\frac{a\sin^2\theta}{\Xi} \, \df \widetilde{\phi} \bigg) \nnr
& \quad + \frac{Q_1}{\beta^2 W}(a\cos\theta -\beta^{-1}\Delta_{\Delta u}) \bigg( \df t- \frac{L}{\Xi a}  \, \df \widetilde{\phi} \bigg) ,\nnr
\widetilde{A}_{2} & = \frac{P^2}{\beta^2 W}( r + \beta^{-1}\Delta_{\Delta r}) \bigg( \df t -\frac{a\sin^2\theta}{\Xi} \, \df \widetilde{\phi} \bigg) \nnr
& \quad + \frac{Q_2}{\beta^2 W}(a\cos\theta +\beta^{-1}\Delta_{\Delta u}) \bigg( \df t- \frac{L}{\Xi a} \, \df \widetilde{\phi} \bigg) ,
\end{align}
where $W(r, \gq)$ is given by \eq{defW} and $L(r)$ is given by \eq{defL}.


\subsection{Known subcases}



\subsubsection{Kerr--Newman--AdS}


When $\delta_1 = \delta_2$, $\gamma_1 = \gamma_2$ and $N = 0$, one recovers the dyonic Kerr--Newman--AdS black hole \cite{Carter, Carter:1968ks}. We provide some details of this solution, showing its embedding in our more general solution, since it might be the case best known to the reader.  The scalar fields vanish, $\chi = 0$, $\varphi = 0$ and the gauge fields are equal, $A_1=A_2$, $\widetilde A^1 = \widetilde A^2$. The field equations are the Einstein--Maxwell equations with a cosmological constant.  We have $v=0$, $ \beta^4 e^2=(Q_1)^2+(P^1)^2$ and the solution is the standard Kerr--Newman--AdS solution (see e.g.\ \cite{Caldarelli:1999xj}).  To match the literature, we use the rescaled charge parameters $\overline{q} = Q_1/\beta^2$ and $\overline{p} = P^1/\beta^2$.  Then the metric is
\begin{align}
\ds & = -\frac{R_g}{W}\left( \df t-\frac{a}{\Xi} \sin^2\theta \, \df \widetilde{\phi} \right)^2 + W  \bigg( \fr{\df r^2}{R_g} + \fr{\df \theta^2}{\Theta_g} \bigg) \nnr
& \quad + \frac{\Theta_g \, \sin^2\theta}{W }\left(a \, \df t - \frac{r^2+a^2}{\Xi} \, \df \widetilde{\phi} \right)^2 \label{AKN}
\end{align}
with
\begin{align}
W(r,\theta) &= r^2 + a^2\cos^2\theta , \nnr
R_g(r) & =  (1+ g^2r^2) ( r^2 + a^2 )-2 m r + \overline{q}^2 + \overline{p}^2 ,\nnr
\Theta_g(\theta) & = 1 - a^2 g^2 \cos^2\theta.
\end{align}
The gauge field and dual gauge field are
\begin{align}
A^{1} & = \frac{\overline{q} r }{W} \bigg( \df t - \frac{a\sin^2\theta}{\Xi} \, \df \widetilde{\phi} \bigg) \nnr
& \quad - \frac{\overline{p} a \cos\theta}{W} \bigg( \df t - \frac{r^2+a^2}{\Xi a} \, \df \widetilde{\phi} \bigg) , \nnr
\widetilde{A}_{1} & = \frac{\overline{p} r}{W} \bigg( \df t -\frac{a\sin^2\theta}{\Xi}\, \df \widetilde{\phi} \bigg) \nnr
& \quad + \frac{\overline{q} a \cos\theta}{W} \bigg( \df t- \frac{r^2+a^2}{\Xi a} \, \df \widetilde{\phi} \bigg) .
\end{align}
The physical electric and magnetic charges will be derived in \eqref{QP}. Note that the parametrization of the electric and magnetic charge is still intricate. One has
\begin{align}
\overline{q} & = \beta^{-2} Q_1 = \beta^{-2}\widehat m \frac{\cosh (4\gamma_1) \, \sinh (2 \delta_1)}{\cosh (2\gamma_1)}, \nnr
\overline{p} & = \beta^{-2} P^1 = \beta^{-2}\widehat m \frac{\sinh (2\gamma_1)}{\cosh (2\delta_1)}. \label{sys}
\end{align}
If however the black hole is only electrically or magnetically charged, then $\beta = 1$ and the parametrization is optimal in terms of a single hyperbolic function.


\subsubsection{Ungauged solutions}


In the ungauged limit $g=0$, one recovers the solution of $\cN = 4$ supergravity \cite{LozanoTellechea:1999my}, restricting to 2 of the 6 vectors, and to scalars that vanish at infinity.  A further specialization of taking $\gd_2 = \gc_2 = 0$ gives the solution of Einstein--Maxwell--dilaton--axion gravity \cite{Galtsov:1994pd}, restricting to scalars that vanish at infinity.  To match with these, take our solution in $(\widehat{t}, \widehat{r}, \widehat{u}, \widehat{\phi})$ coordinates.  Note that
\be
\widehat{W} = (\widehat{r} + \Sigma_{\Delta r})^2 + (\widehat{u} + \Sigma_{\Delta u})^2  - \upsilon^2 ,
\ee
where
\be
(M^2 + N^2) \upsilon^2 = \Delta_2^2 -\Delta_4
\ee
and $\Delta_2$ and $\Delta_4$ are defined as
\begin{align}
\Delta_2 & = \frac{1}{4}\sum_{a=1}^2 \big( (Q_a)^2+(P^a)^2 \big), & \Delta_4 & = \frac{1}{4} (Q_1 Q_2 + P^1 P^2)^2.
\end{align}
Alternatively, we have
\begin{align}
16 (M^2 + N^2) \upsilon^2 & = [(Q_1)^2 + (P^2)^2 - (Q_2)^2 - (P^1)^2]^2 \nnr
& \quad + 4 (Q_1 P^1 - Q_2 P^2)^2 ,
\end{align}
which shows that $\upsilon^2$ is non-negative.  To match with \cite{LozanoTellechea:1999my}, we dualize the second gauge field and make the identifications $Q_1 = 2 Q^{(1)}$, $P_1 = 2 P^{(1)}$, $Q_2 = 2 P^{(2)}$, $P^2 = - 2 Q^{(2)}$, so that $\upsilon^2 = | \Upsilon |^2$ there.  Now make the coordinate change $\widehat{r} + \Sigma_{\Delta r} = r + M$ and $\widehat{u} + \Sigma_{\Delta u} = N + \alpha \cos \gq$, $\widehat{a} = \alpha$, and the solution is seen to match.  To match with \cite{Galtsov:1994pd}, set $\gd_2 = \gc_2 = 0$, $Q_1 = 2 Q$, $P^1 = 2 P$, and perform similar translations of $\widehat{r}$ and $\widehat{u}$.

This ungauged solution is also a subset of the general black hole solution discussed in \cite{Chow:2013tia}, which has 4 electric charges and 4 magnetic charges. Indeed, defining the shifted coordinate $\widehat t_0$ as
\be
\widehat t = \widehat t_0 - \frac{\Delta \widehat u_1 \Delta \widehat u_2}{\widehat a}\widehat \phi,
\ee
the metric reads as
\begin{align}
\ds & = \widehat W \bigg( \fr{\df \widehat r^2}{\widehat R_g} + \fr{\df \widehat u^2}{\widehat U_g} \bigg) - \fr{\widehat R_g}{\widehat W} \bigg( \df \widehat  t_0 - \fr{\widehat U(\widehat u) - 2 N \widehat u}{\widehat a} \, \df \widehat \gf \bigg)^2  \nnr
& \quad + \fr{\widehat U_g}{\widehat W} \bigg( \df\widehat   t_0 - \fr{\widehat R(\widehat r) + 2 M \widehat r + V_2}{\widehat a} \, \df \widehat \gf \bigg)^2 ,
\end{align}
where
\begin{align}
\widehat W(\widehat r,\widehat u) & = R(\widehat r)- U(\widehat u) + 2 M \widehat r + 2 N \widehat u + V_2,\nnr
\widehat R_g(\widehat r) & = R(\widehat r) + g^2 \widehat r_1 \widehat r_2 (\widehat r_1 \widehat r_2 + \widehat a^2),\nnr
\widehat U_g(\widehat u)& = U(\widehat u) + g^2 \widehat u_1 \widehat u_2 (\widehat u_1 \widehat u_2 - \widehat a^2),\nnr
R(\widehat r) & = \widehat r^2 - 2 \widehat m \widehat r + \widehat a^2 \nnr
U(\widehat u) & = -\widehat u^2 + 2 \widehat n \widehat u + \widehat a^2,\nnr
V_2 & = \Delta \widehat r_1 \Delta \widehat r_2 + \Delta \widehat u_1 \Delta \widehat u_2.
\end{align}
One can compare this metric to the one written in \cite{Chow:2013tia}, when one restricts the solution of \cite{Chow:2013tia} to pairwise equal gauge fields in order to be a solution of $\cN= 2$, $\textrm{U}(1)^2$ supergravity. The metrics match, upon recognizing that the function $L(r)$ defined in \cite{Chow:2013tia} is given by $2 M r+V_2$ in the case of pairwise equal gauge fields, and making the change of parameters and variables $\widehat{a}^2 = a^2 - n^2$ and $\widehat{\phi}/\widehat{a} = \phi/a$.


\subsubsection{Gauged solutions}


If $\gc_1 = \gc_2 = 0$, then we have the solution of \cite{Chong:2004na} that is electrically charged and includes a NUT charge.  If furthermore the NUT charge vanishes, then we have the simplifications that $\Delta \widehat u_1 = \Delta \widehat u_2 = 0$, $\widehat a = a$ and $\beta = 1$.  We shall later reexamine the supersymmetric asymptotically AdS solutions within this class.


\subsection{Thermodynamics}
\label{thermo2}



\subsubsection{Conserved charges}
\label{thermo2mass}


The crux of the analysis of black hole thermodynamics is the definition of the black hole mass. We follow here the canonical Lagrangian definitions \cite{Regge:1974zd,Abbott:1981ff,Iyer:1994ys,Barnich:2001jy,Barnich:2007bf}. The explicit form of the conserved charge associated with a symmetry of a Lagrangian of the form \eqref{generalaction} has been derived in \cite{Compere:2009dp}.  The definition of angular momentum can be obtained from the same formalism.  For special cases, the thermodynamical quantities have been previously computed in \cite{Kostelecky:1995ei, Caldarelli:1999xj, Cvetic:2005zi,Papadimitriou:2005ii}.

We will set the NUT charge to zero from now on, which implies $n=N=N_g=0$.  As emphasized in \cite{Gibbons:2004ai}, the thermodynamics is best understood using a non-rotating frame at infinity.  Since the coordinate frame $(t, r, \theta, \phi)$ is non-rotating at infinity, in these coordinates the mass and angular momentum and associated with the respective Killing vectors $\pd/\pd t$ and $- \pd/\pd \phi$.

The gravitational contribution (which is derived from the Einstein action) to the angular momentum is finite and integrable. The gravitational contribution to the mass is linearly divergent in $r$ and not integrable. The matter sector therefore contributes to the charges. Looking at the asymptotic form of the canonical expression for the conserved charges, we notice that the gauge fields do not contribute to the mass or angular momentum. The scalar fields do not contribute to the angular momentum but they do contribute to the mass. More precisely, the scalar fields contribute to a linearly divergent piece in $r$, but do not contribute to the finite $r^0$ piece. This linearly divergent piece exactly cancels the divergence in the gravitational contribution, after using the explicit asymptotic form
\begin{align}
\varphi & = 2\frac{\Delta_{\Delta r}}{\beta r} + \frac{2\Delta_{\Delta u}(a \beta  \cos\theta + \Delta_{\Delta u})}{\beta^2 r^2}+O(r^{-3}), \nnr
\chi & = 2\frac{\Delta_{\Delta u}}{\beta r} - \frac{2 \Delta_{\Delta r}(a \beta \cos\theta + 2 \Delta_{\Delta u})}{\beta^2 r^2}+O(r^{-3}),
\end{align}
and the expression for $v$ in \eqref{vv}. The conserved mass is then finite and integrable.

The final expressions for the mass and angular momentum are
\begin{align}
M & = \frac{m}{G \Xi^2} , & J & = \frac{m a}{G \Xi^2} .
\end{align}
These expressions are familiar already for the Kerr--AdS black hole \cite{Gibbons:2004ai,Deruelle:2004mv,Barnich:2004uw}. Since the matter fields do not contribute to the finite conserved charges, it is not surprising that the mass and angular momentum agree with the uncharged black hole.

Using the canonical expressions for the charges derived from the action, one also obtains that the electromagnetic charges in geometrical units are
\begin{align}
\overline{Q}_a & = \frac{Q_a}{2G \beta^2 \Xi}, & \overline{P}^a & = \frac{P^a}{2G \beta^2 \Xi} .\label{QP}
\end{align}
The factor of 2 difference between the normalization of the gauge kinetic terms between \eqref{Lthird} and \eqref{LU2} is responsible for the factor of 2 between the definitions of $\overline{Q}_a,\, \overline{P}^a$ defined here and $\overline{Q}_I,\, \overline{P}^I$, $I=1,2,3,4$ defined in Section \ref{consch}. The factor of $1/\Xi$ is familiar already for the Kerr--Newman--AdS black hole \cite{Kostelecky:1995ei, Caldarelli:1999xj}. The factor of $\beta^{-2}$ is a new feature for dyonic Kerr--Newman--AdS black holes in our parametrization.


\subsubsection{First law}
\label{fl2}


The inner and outer black hole horizons are located at $r=r_\pm$ which are the zeros of the radial function $R_g(r)$ defined in \eqref{Rg}.  The Killing generator is $\xi^\mu \, \pd_\mu = \pd_t + \Omega \, \pd_\phi$, where the angular velocity is
\be
\Omega = a \bigg( \frac{\Xi}{L(r_+)} + g^2 \bigg) .
\ee
The temperature and entropy of the outer horizon are
\begin{align}
T & = \frac{R_g'(r_+)}{4\pi L(r_+)}, & S & = \frac{\pi L(r_+)}{\Xi \, G},
\end{align}
where $L$ is defined in \eqref{defL}. The electric and magnetic potentials are defined as the difference between the potentials at the horizon and at infinity,
\begin{align}
\Phi^a & = \xi^\mu A^a_{\mu}|_{r=r_+} - \xi^\mu A^a_{\mu}|_{r=\infty}, \nnr
\Psi_a & = \xi^\mu \widetilde{A}_{a \mu}|_{r=r_+}-\xi^\mu \widetilde{A}_{a \mu}|_{r=\infty} .
\end{align}
After remarkable simplifications, one obtains
\begin{align}
\Phi^1 & = \frac{Q_1 (\beta r_+ + \Delta_{\Delta r}) - P^1 \Delta_{\Delta u}}{\beta^3 L(r_+)}, \nnr
\Phi^2 & = \frac{Q_2 (\beta r_+ - \Delta_{\Delta r}) + P^2 \Delta_{\Delta u}}{\beta^3 L(r_+)}, \nnr
\Psi_1 & = \frac{P^1 (\beta r_+ - \Delta_{\Delta r})-Q_1 \Delta_{\Delta u}}{\beta^3 L(r_+)}, \nnr
\Psi_2 & = \frac{P^2 (\beta r_+ + \Delta_{\Delta r})+Q_2 \Delta_{\Delta u}}{\beta^3 L(r_+)} .
\end{align}
The first law
\be
\delta M = T \, \delta S + \Omega \, \delta J + \Phi^a \, \delta \overline{Q}_a + \Psi_a \, \delta \overline{P}^a,
\ee
is obeyed for generic variations, which provides with a non-trivial check of our expressions. In the static case, we checked explicitly that the symplectic flux is zero at $r \rightarrow \infty$, see Section \ref{secmass}. Boundary conditions with varying electric and magnetic charges in $\cN= 2$, $\textrm{U}(1)^2$ gauged supergravity are therefore consistent on the restricted phase space of black hole solutions. It was then expected that the first law holds for static configurations. The closure of the first law in the general rotating case suggests that no qualitative change occurs when angular momentum is turned on. In particular, we checked explicitly that the symplectic flux is zero at $r \rightarrow \infty$ for the Kerr--Newman--AdS solutions.


\subsection{Supersymmetric solutions}


For purely electric solutions, the Bogomolny--Prasad--Sommerfield (BPS) condition for supersymmetric solutions was given in \cite{Cvetic:2005zi}.
\be
M = gJ + \overline{Q}_1 + \overline{Q}_2 ,
\ee
after choosing signs so that $g$, $J$, $\overline{Q}_1$ and $\overline{Q}_2$ are non-negative.  The subsequent analysis was been performed previously, however to correct a previous typographical error we repeat the calculation.  The BPS condition is
\be
\expe{2 (\gd_1 + \gd_2)} = 1 + \frac{2}{a g} .
\ee
With this condition, we have
\begin{align}
R_g & = g^2 \bigg( r_1 r_2 - \fr{2}{g^2 (\expe{2 (\gd_1 + \gd_2)} - 1)} \bigg) ^2 \nnr
& \quad + \coth^2 (\gd_1 + \gd_2) \bigg( r - \fr{2 m s_{\gd 1} s_{\gd 2}}{\cosh (\gd_1 + \gd_2)} \bigg) ^2 .
\end{align}
This is a sum of two squares, so at a horizon both squares must vanish.  In general, the zeros of the two squares are different, so the supersymmetry condition alone leads to a solution that is singular.  If there is a horizon, then the second square implies that it is at
\be
r = r_0 \equiv \fr{2 m s_{\gd 1} s_{\gd 2}}{\cosh (\gd_1 + \gd_2)} .
\ee
Substituting into the first square and requiring it to vanish gives an additional condition has to be imposed so that the solution will have a regular horizon, namely
\be
m^2 g^2 = \fr{\cosh^2 (\gd_1 + \gd_2)}{\expe{\gd_1 + \gd_2} \sinh^3 (\gd_1 + \gd_2) \sinh(2 \gd_1) \sinh(2 \gd_2)} .
\ee
$R_g$ then possesses a double root at $r = r_0$, which indicates that the temperature vanishes, as it must for a supersymmetric solution.  The supersymmetric Kerr--Newman--AdS black hole \cite{Kostelecky:1995ei, Caldarelli:1998hg} is recovered when furthermore $\gd_1 = \gd_2$.

There are complications with BPS bounds when magnetic charge is included.  For example, in minimal gauged supergravity two different BPS bounds were found \cite{Hristov:2011ye}, corresponding to different superalgebras.  The analysis has been generalized to include more general matter couplings \cite{Hristov:2011qr}.

An alternative, and definitive, approach to finding supersymmetric solutions is to find Killing spinors.  This approach has recently been carried out in full \cite{Klemm:2013eca} for the Pleba\'{n}ski--Demia\'{n}ski solution \cite{Plebanski:1976gy}, which includes the dyonic Kerr--Newman--Taub--NUT--AdS solution, of Einstein--Maxwell theory with a negative cosmological constant, i.e.\ the bosonic sector of minimal $\cN = 2$ gauged supergravity.  Within the Kerr--Newman--AdS family, supersymmetric black holes must carry only electric charge \cite{Caldarelli:1998hg}.  Supersymmetric solutions of $\cN = 2$ gauged supergravity coupled to vector multiplets have been classified \cite{Cacciatori:2008ek, Klemm:2009uw}.


\subsection{Killing tensors and separability}


The spacetime admits a web of various interrelated (conformal) Killing tensors.  For reviews of these tensors, see e.g.\ \cite{Yasui:2011pr}.  They are related to the separability of equations for geodesic motion and for probe scalar fields.


\subsubsection{Killing tensors}


We now recall some relevant defintions of Killing tensors.  A Killing--St\"{a}ckel (KS) tensor $K_{\mu \nu} = K_{(\mu \nu)}$ satisfies $\nabla_{(\mu} K_{\nu \rho)} = 0$.  A conformal Killing--St\"{a}ckel (CKS) tensor $Q_{\mu \nu} = Q_{(\mu \nu)}$ satisfies $\nabla_{(\mu} Q_{\nu \rho)} = q_{(\mu} g_{\nu \rho)}$ for some $q_\mu$, given in 4 dimensions by $q_\mu = \tf{1}{6} (\pd_\mu Q{^\nu}{_\nu} + 2 \nabla_\nu Q{^\nu}{_\mu})$.  A Killing--Yano (KY) tensor $Y_{\mu \nu} = Y_{[\mu \nu]}$ satisfies $\nabla_{(\mu} Y_{\nu) \rho} = 0$.  A Killing--Yano tensor with torsion (KYT tensor), $Y_{\mu \nu} = Y_{[\mu \nu]}$, satisfies $\nabla^T_{(\mu} Y_{\nu) \rho} = 0$, where the covariant derivative $\nabla^T$ uses the connection $\gC{^\mu}{_{\nu \rho}} + T{^\mu}{_{\nu \rho}}$, including both the Levi-Civita connection $\gC{^\mu}{_{\nu \rho}}$ and a torsion $T{^\mu}{_{\nu \rho}}$ such that $T_{\mu \nu \rho} = T_{[\mu \nu \rho]}$, i.e.\ derived from a 3-form.  A conformal Killing--Yano tensor with torsion (CKYT tensor), $k_{\mu \nu} = k_{[\mu \nu]}$, satisfies $\nabla^T_{(\mu} k_{\nu) \rho} = k_\rho g_{\mu \nu} - k_{(\mu} g_{\nu) \rho}$ for some $k_\mu$, given in 4 dimensions by $k_\mu = \tf{1}{3} \nabla{^T}{_\nu} k{^\nu}{_\mu}$.  A CKYT tensor is a closed conformal Killing--Yano tensor with torsion (CCKYT tensor) if furthermore $\nabla{^T}{_{[\mu}} k_{\nu \rho]} = 0$.  The literature sometimes uses ``generalized'' to mean ``with torsion''.

We now recall some relevant results about Killing tensors.  The Hodge dual of a KYT tensor is a CCKYT tensor (with the same torsion), and vice versa.  If $Y_{\mu \nu}$ is a KY(T) tensor, then $K_{\mu \nu} = Y{_\mu}{^\rho} Y_{\rho \nu}$ is a KS tensor.  If $Q^{\mu \nu}$ is a CKS tensor, then the components $Q^{\mu \nu}$ give a CKS tensor for any conformally related metric.

There are two metrics of interest: the usual Einstein frame metric $\df s^2$, and the conformally related string frame metric
\be
\df \wtd{s}^2 = \fr{r^2 + u^2}{W} \, \df s^2 .
\label{stringframemetric}
\ee
For black hole solutions of supergravity, usually only the string frame metric admits a KS tensor \cite{{Chow:2008fe}}, not the Einstein frame metric.  However, for the special solutions considered here, both the string frame and Einstein frame metrics admit KS tensors.  Both KS tensors have two ``square roots'' given by KYT tensors.

If we make the coordinate change $\tau = \widehat t - a \widehat \phi$, $\psi = \widehat \phi/a$, then the Einstein frame metric \eqref{mads3} is
\begin{align}
\ds & = - \fr{R_g}{W} (\df \tau + u_1 u_2 \, \df \psi)^2 + \fr{W}{R_g} \, \df r^2 \nnr
& \quad + \fr{U_g}{W} (\df \tau - r_1 r_2 \, \df \psi)^2 + \fr{W}{U_g} \, \df u^2 .
\end{align}
Here and below, we omit the hat on functions and coordinates for simplicity. The gauge fields are
\begin{align}
A^1 & = \fr{1}{ W} \bigg(Q_1 r_2 (\df \tau + u_1 u_2 \, \df \psi) -P^1 u_2 (\df \tau - r_1 r_2 \, \df \psi) \bigg) , \nnr
A^2 & = \fr{1}{ W} \bigg( Q_2 r_1 (\df \tau + u_1 u_2 \, \df \psi) -P^2 u_1 (\df \tau - r_1 r_2 \, \df \psi) \bigg) ,
\end{align}
and the dual gauge fields are
\begin{align}
\wtd{A}_1 & = \fr{1}{ W} \bigg( P^1 r_1 (\df \tau + u_1 u_2 \, \df \psi)  + Q_1 u_1 (\df \tau - r_1 r_2 \, \df \psi) \bigg) , \nnr
\wtd{A}_2 & = \fr{1}{ W} \bigg( P^2 r_2 (\df \tau + u_1 u_2 \, \df \psi)  +Q_2 u_2 (\df \tau - r_1 r_2 \, \df \psi) \bigg) . \end{align}
In fact, these are probably the simplest coordinates for expressing the solution locally.

Consider more generally the metric
\begin{align}
\ds & = - \fr{R}{W} (\df \tau + W_u \, \df \psi)^2 + \fr{W}{R} \, \df r^2 \nnr
& \quad + \fr{U}{W} (\df \tau - W_r \, \df \psi)^2 + \fr{W}{U} \, \df u^2 ,
\end{align}
where $W = W_r + W_u$; $R$ and $W_r$ are arbitrary functions of $r$; and $U$ and $W_u$ are arbitrary functions of $u$. Here and below we drop the subscript $_g$ in $R$ and $U$ for simplicity. Introduce the vielbeins
\begin{align}
e^0 & = \fr{\sr{R}}{\sr{W}} (\df \tau + W_u \, \df \psi) , & e^1 & = \fr{\sr{W}}{\sr{R}} \, \df r , \nnr
e^2 & = \fr{\sr{U}}{\sr{W}} (\df \tau - W_r \, \df \psi) , & e^3 & = \fr{\sr{W}}{\sr{U}} \, \df u .
\end{align}
Two KYT tensors are
\be
Y_\pm = \sr{W_u} e^0 \wedge e^1 \pm \sr{W_r} e^2 \wedge e^3 ,
\ee
with corresponding torsions
\begin{align}
T_\pm & = \fr{1}{W} \bigg[ U \bigg( \pd_r W_r \mp \fr{\sr{W_r}}{\sr{W_u}} \pd_u W_u \bigg) \, \df r \nnr
& \quad + R \bigg( \pd_u W_u \mp \fr{\sr{W_u}}{\sr{W_r}} \pd_r W_r \bigg) \, \df u \bigg] \wedge \df \tau \wedge \df \psi .
\end{align}
Squaring either $Y_\pm$ gives the KS tensor
\be
K_{\mu \nu} \, \df x^\mu \, \df x^\nu = W_u (- e^0 e^0 + e^1 e^1) - W_r (e^2 e^2 + e^3 e^3) .
\ee
Taking Hodge duals of $Y_\pm$ gives the CCKYT tensors
\be
k_\pm = \pm \sr{W_r} e^0 \wedge e^1 - \sr{W_u} e^2 \wedge e^3 .
\ee
If we make the coordinate change $r' = \sr{W_r}$, $u' = \sr{W_u}$, and define the functions $f_r = 1/ (\pd_r \sr{W_r})$, $f_u = 1/ (\pd_u \sr{W_u})$, $R' = R/f_r^2$, $U' = U/f_u^2$, then the metric takes the form
\begin{align}
\ds & = - \fr{R' f_r^2}{r'{^2} + u'{^2}} (\df \tau + u'{^2} \, \df \psi)^2 + \fr{r'{^2} + u'{^2}}{R'} \, \df r'{^2}  \nnr
& \quad + \fr{U' f_u^2}{r'{^2} + u'{^2}} (\df \tau - r'{^2} \, \df \psi)^2 + \fr{r'{^2} + u'{^2}}{U'} \, \df u'{^2} .
\end{align}
The torsions take the form
\begin{align}
T_\pm & = - \fr{2}{r'{^2} + u'{^2}} \bigg[ \bigg( \fr{f_u}{f_r} \mp 1 \bigg) \fr{r' \sr{U'}}{\sr{r'{^2} + u'{^2}}} e^1 \nnr
& \quad + \bigg( \fr{f_r}{f_u} \mp 1 \bigg) \fr{u' \sr{R'}}{\sr{r'{^2} + u'{^2}}} e^3 \bigg] \wedge e^0 \wedge e^2 .
\end{align}
These forms of the metric and torsion $T_+$ manifestly fit into the classification of metrics admitting a KYT tensor \cite{Houri:2012eq}, specifically even-dimensional of type A, after analytically continuing to Riemannian signature.

The string frame metric \eq{stringframemetric} can be expressed in terms of the vielbeins
\begin{align}
\wtd{e}^0 & = \fr{\sr{(r^2 + u^2) R}}{W} (\df \tau + W_u \, \df \psi) , & \wtd{e}^1 & = \fr{\sr{r^2 + u^2}}{\sr{R}} \, \df r , \nnr
\wtd{e}^2 & = \fr{\sr{(r^2 + u^2) U}}{W} (\df \tau - W_r \, \df \psi) , & \wtd{e}^3 & = \fr{\sr{r^2 + u^2}}{\sr{U}} \, \df u .
\end{align}
Two KYT tensors are
\be
\wtd{Y}_\pm = u \wtd{e}^0 \wedge \wtd{e}^1 \pm r \wtd{e}^2 \wedge \wtd{e}^3 ,
\ee
with corresponding torsions
\begin{align}
\wtd{T}_\pm & = \fr{r^2 + u^2}{W} \bigg[ U \bigg( \fr{\pd_r W_r}{W} \mp \fr{2 r}{r^2 + u^2}\bigg) \, \df r \nnr
& \quad + R \bigg( \fr{\pd_u W_u}{W} \mp \fr{2 u}{r^2 + u^2} \bigg) \, \df u \bigg] \wedge \df \tau \wedge \df \psi \nnr
& = - \bigg[ \bigg( \fr{\pd_r W_r}{W} \mp \fr{2 r}{r^2 + u^2}\bigg) \fr{\sr{U}}{\sr{r^2 + u^2}} \wtd{e}^1 \nnr
& \quad + \bigg( \fr{\pd_u W_u}{W} \mp \fr{2 u}{r^2 + u^2} \bigg) \fr{\sr{R}}{\sr{r^2 + u^2}} \wtd{e}^3 \bigg] \wedge \wtd{e}^0 \wedge \wtd{e}^2 .
\end{align}
Squaring either $\wtd{Y}_\pm$ gives the KS tensor
\be
\wtd{K}_{\mu \nu} \, \df x^\mu \, \df x^\nu = u^2 (- \wtd{e}^0 \wtd{e}^0 + \wtd{e}^1 \wtd{e}^1) - r^2 (\wtd{e}^2 \wtd{e}^2 + \wtd{e}^3 \wtd{e}^3) .
\ee
Taking Hodge duals of $\wtd{Y}_\pm$ gives the CCKYT tensors
\be
\wtd{k}_\pm = \pm r \wtd{e}^0 \wedge \wtd{e}^1 - u \wtd{e}^2 \wedge \wtd{e}^3 .
\ee
The string frame metric can also be written as
\begin{align}
\df \wtd{s}^2 & = - \fr{R}{r^2 + u^2} (\df \tau + u^2 \, \df \psi - \cA)^2 + \fr{r^2 + u^2}{R} \, \df r^2 \nnr
& \quad + \fr{U}{r^2 + u^2} (\df \tau - r^2 \, \df \psi - \cA)^2 + \fr{r^2 + u^2}{U} \, \df u^2 ,
\end{align}
where
\begin{align}
\cA & = \fr{r^2 + u^2}{W} \bigg( \fr{W_r - r^2}{r^2 + u^2} (\df \tau + u^2 \, \df \psi) \nnr
& \quad  + \fr{W_u - u^2}{r^2 + u^2} (\df \tau - r^2 \, \df \psi)\bigg) .
\end{align}
The torsion $\wtd{T}_+$ can also be written as
\begin{align}
\wtd{T}_+ & = - \bigg[ \pd_r \log \bigg( \fr{W}{r^2 + u^2} \bigg) \fr{\sr{U}}{\sr{r^2 + u^2}} \wtd{e}^1 \nnr
& \quad + \pd_u \log \bigg( \fr{W}{r^2 + u^2} \bigg) \fr{\sr{R}}{\sr{r^2 + u^2}} \wtd{e}^3 \bigg] \wedge \wtd{e}^0 \wedge \wtd{e}^2 .
\end{align}
These forms of the metric and torsion $T_+$ also manifestly fit into the classification of metrics admitting a KYT tensor \cite{Houri:2012eq}, again even-dimensional of type A, after analytically continuing to Riemannian signature.

The string frame KS tensor $\wtd{K}_{\mu \nu}$ induces a CKS tensor $Q_{\mu \nu}$ for the Einstein frame metric, with components $Q^{\mu \nu} = \wtd{K}^{\mu \nu}$.  Similarly, the Einstein frame KS tensor $K_{\mu \nu}$ induces a CKS tensor $\wtd{Q}_{\mu \nu}$ for the string frame metric, with components $\wtd{Q}^{\mu \nu} = K^{\mu \nu}$.  In fact,
\begin{align}
Q_{\mu \nu} & = K_{\mu \nu} + q g_{\mu \nu} , & \wtd{Q}_{\mu \nu} = \wtd{K}_{\mu \nu} + \wtd{q} \wtd{g}_{\mu \nu} ,
\end{align}
where
\begin{align}
q & = \fr{u^2 W_r - r^2 W_u}{r^2 + u^2} , & \wtd{q} & = \fr{r^2 W_u - u^2 W_r}{W} .
\end{align}
Therefore, $\nabla_{(\mu} Q_{\nu \rho)} = q_{(\mu} g_{\nu \rho)}$ and $\nabla_{(\mu} \wtd{Q}_{\nu \rho)} = \wtd{q}_{(\mu} \wtd{g}_{\nu \rho)}$, where $q_\mu = \pd_\mu q$ and $\wtd{q}_\mu = \pd_\mu \wtd{q}$.

Despite the interesting geometrical structures, the physical interpretations of the torsions are generally unclear.  While the torsions $T_+$ and $\wtd{T}_+$ vanish for the uncharged Kerr--Taub--NUT--AdS solution (for which the string and Einstein frames coincide), recovering the known KY tensor, in this limit the torsions $T_-$ and $\wtd{T}_-$ do not vanish.  In special cases, the string frame torsion $\wtd{T}_+$, but not the Einstein frame torsion $T_+$, is physically motivated by dualizing the axion $\chi$ to give a 3-form field strength $H$, which is identified as the torsion \cite{Houri:2010fr}.  Analogously in 5-dimensional minimal gauged supergravity, the vector can be dualized \cite{Kubiznak:2009qi}.  However, the dualization cannot be performed for solutions of gauged supergravity, since the potential \eq{gauged} involves the bare axion potential, not its derivative.  Furthermore, although \cite{Houri:2010fr} found the correct torsion by dualizing the axion for the ``Kerr--Sen'' black hole of ungauged supergravity (the $\gd_2 = \gc_1 = \gc_2 = \widehat{n} = g = 0$ solution), for which there is a single electric gauge field, the procedure fails for our more general dyonic solution.  More specifically, in the Kerr--Sen case $\df \wtd{T}_+ + F^1 \wedge F^1 = 0$, which is consistent with identifying the torsion with $H$, but $\df \wtd{T}_+ + F^1 \wedge F^1 \neq 0$ when we generalize to $\gc_1 \neq 0$.


\subsubsection{Separability}


The KS tensors in Einstein frame and string frame guarantee the complete integrability of geodesic motion in both these frames, which we now demonstrate explicitly.  The Einstein and string frame metric have respective inverses
\begin{align}
\bigg( \fr{\pd}{\pd s} \bigg) ^2 & = \fr{1}{W} \bigg( - \fr{(W_r \, \pd_\tau + \pd_\psi)^2}{R} + R \, \pd_r^2 \nnr
& \quad + \fr{(W_u \, \pd_\tau - \pd_\psi)^2}{U} + U \, \pd_u^2 \bigg) , \nnr
\bigg( \fr{\pd}{\pd \wtd{s}} \bigg) ^2 & = \fr{1}{r^2 + u^2} \bigg( - \fr{(W_r \, \pd_\tau + \pd_\psi)^2}{R} + R \, \pd_r^2 \nnr
& \quad + \fr{(W_u \, \pd_\tau - \pd_\psi)^2}{U} + U \, \pd_u^2 \bigg) ,
\end{align}
and metric determinants given by
\begin{align}
\sr{-g} & = W , & \sr{- \wtd{g}} & = \fr{(r^2 + u^2)^2}{W} .
\end{align}

In Einstein frame, the Hamilton--Jacobi equation for geodesic motion is
\be
\fr{\pd S}{\pd \lambda} + \fr{1}{2} g^{\mu \nu} \, \pd_\mu S \, \pd_\nu S = 0 ,
\ee
where $S$ is Hamilton's principal function, $\pd_\mu S = p_\mu = \df x_\mu / \df \gl$, $p_\gl$ are momenta conjugate to $x^\mu$, and $\gl$ is an affine parameter.  Consider the ansatz
\be
S = \tf{1}{2} \mu^2 \gl - E \tau + L \psi + S_r (r) + S_u (u) .
\ee
The constants $p_\tau = - E$ and $p_\phi = L$ are momenta conjugate to the ignorable coordinates $\tau$ and $\psi$, related to energy and angular momentum.  The particle mass is $\mu$, so that $p^\mu p_\mu = - \mu^2$.  The components $W g^{\mu \nu}$ are additively separable into functions of $r$ and of $u$, and so the Hamilton--Jacobi equation is additively separable.  Explicitly, we have
\begin{align}
& - \fr{(W_r E - L)^2}{R} + \fr{(W_u E + L)^2}{U} + R \bigg( \fr{\df S_r}{\df r} \bigg) ^2 \nnr
& + U \bigg( \fr{\df S_u}{\df u} \bigg) ^2 + \mu^2 (W_r + W_u) = 0 ,
\label{KleinGordon}
\end{align}
and so
\begin{align}
\fr{\df S_r}{\df r} & = \fr{1}{R} \sr{(W_r E - L)^2 - (C + \mu^2 W_r) R} , \nnr
\fr{\df S_u}{\df u} & = \fr{1}{U} \sr{- (W_u E + L)^2 + (C- \mu^2 W_u) U} ,
\end{align}
where $C$ is a separation constant.  We then determine $r(\gl)$ and $u (\gl)$ by integrating
\begin{align}
\fr{\df r}{\df \gl} & = g^{r r} p_r = \fr{R}{W} \fr{\df S_r}{\df r} , & \fr{\df u}{\df \gl} & = g^{u u} p_u = \fr{U}{W} \fr{\df S_u}{\df u} .
\end{align}
Finally, we determine $\tau (\gl)$ and $\psi (\gl)$ by integrating
\begin{align}
\fr{\df \tau}{\df \gl} & = g^{\tau \tau} p_\tau + g^{\tau \psi} p_\psi \nnr
& = \fr{E}{W} \bigg( \fr{W_r^2}{R} - \fr{W_u^2}{U} \bigg) - \fr{L}{W} \bigg( \fr{W_r}{R} + \fr{W_u}{U} \bigg) , \nnr
\fr{\df \psi}{\df \gl} & = g^{\tau \psi} p_\tau + g^{\psi \psi} p_\psi \nnr
& = \fr{E}{W} \bigg( \fr{W_r}{R} + \fr{W_u}{U} \bigg) + \fr{L}{W} \bigg( \fr{1}{U} - \fr{1}{R} \bigg) .
\end{align}
The separation can analogously be explicitly demonstrated in string frame, by replacing $\mu^2 (W_r + W_u)$ in \eq{KleinGordon} by $\mu^2 (r^2 + u^2)$.

The massive Klein--Gordon equation for the Einstein frame metric is
\be
\square \Phi = \fr{1}{\sr{-g}} \pd_\mu (\sr{-g} g^{\mu \nu} \pd_\nu \Phi) = \mu^2 \Phi .
\ee
Consider the ansatz
\be
\Phi = \Phi_r (r) \Phi_u (u) \expe{\im (k \psi - \omega \tau)} .
\ee
Then the Klein--Gordon equation gives
\begin{align}
\mu^2 W & = \fr{(\omega W_r - k)^2}{R} - \fr{(\omega W_u + k)^2}{U} + \fr{1}{\Phi_r} \fr{\df}{\df r} \bigg( R \fr{\df \Phi_r}{\df r} \bigg) \nnr
& \quad + \fr{1}{\Phi_u} \fr{\df}{\df u} \bigg( U \fr{\df \Phi_u}{\df u} \bigg) ,
\end{align}
which separates to give
\begin{align}
& \fr{\df}{\df r} \bigg( R \fr{\df \Phi_r}{\df r} \bigg) + \bigg( \fr{(\omega W_r - k)^2}{R} - \mu^2 W_r + C \bigg) \Phi_r = 0 , \nnr
& \fr{\df}{\df u} \bigg( U \fr{\df \Phi_u}{\df u} \bigg) - \bigg( \fr{(\omega W_u + k)^2}{U} + \mu^2 W_u + C \bigg) \Phi_u = 0 ,
\end{align}
where $C$ is an integration constant.  For the black hole solutions we found, these are Fuchsian second-order ordinary differential equations.

For other examples \cite{Houri:2010fr, Wu:2009cn, Wu:2009ug}, the existence of KYT tensors is related to the separability of Dirac equations that are modified by terms involving the torsion.  We expect the same separability properties for the class of metrics that we have discussed.


\section{Conclusion}


We extended the known classes of 4-dimensional dyonic AdS black holes in maximal gauged supergravity along two fronts. We obtained a class of static AdS black holes with 4 independent electric charges and 4 independent magnetic charges, and a class of rotating AdS black holes with 2 independent electric and 2 independent magnetic charges (and also an independent NUT charge). In particular, the class of planar dyonic black holes that we derived might be of interest for the AdS/CFT correspondence as models for large $N$ gauge theories with finite charge density and background magnetic field. It is remarkable that such solutions can be obtained from their asymptotically flat cousins by simply modifying one or two functions in the metric while leaving the matter fields unchanged. A natural continuation of this work would be to construct a rotating solution with 8 independent electromagnetic charges, which would require an ansatz that remains to be guessed.  It would be interesting to investigate further the supersymmetry of these solutions.

We showed that varying independently the electric and magnetic charges of dyonic black holes in AdS is in general inconsistent with the existence of a Hamiltonian. Several exceptions exist, including the dyonic Kerr--Newman--AdS family, where the variation of both electromagnetic charges can be performed at the expense of imposing Lorentz-violating boundary conditions for the gauge field. Non-relativistic holographic theories corresponding to such boundary conditions, if they exist, remain to be constructed. More generally, we found several distinct classes of boundary conditions for gauge fields in AdS$_4$. It would be interesting to generalize them to include propagating fields, compute their asymptotic symmetry group and classify their supersymmetric extensions.

Like the Kerr solution, the rotating solutions that we constructed have special algebraic properties, in particular various types of Killing tensors.  We were led to consider a wider class of metrics and found, in two different conformal frames, Killing--Yano tensors for connections with torsion.  Although the physical significance of these torsions is unclear, they underlie the separability of the Hamilton--Jacobi equation for geodesic motion.

\vspace*{10pt}
\begin{center}
\small\textbf{ACKNOWLEDGEMENTS}
\end{center}
\vspace*{10pt}

We gratefully thank the Centro de Ciencias de Benasque Pedro Pascual for its warm hospitality. The work of D.C. was partially supported by the ERC Advanced Grant ``SyDuGraM'', by IISN-Belgium (convention 4.4514.08) and by the ``Communaut\'e Fran\c{c}aise de Belgique" through the ARC program.  G.C. is a Research Associate of the Fonds de la Recherche Scientifique F.R.S.-FNRS (Belgium) and is supported by NSF grant 1205550.

\appendix


\section{Details of the static solution}


Here are the remaining details of the static solution with a spherical horizon.  We denote $s_{\gd I} = \sinh \gd_I$, $c_{\gd I} = \cosh \gd_I$, $s_{\gd I \ldots J} = s_{\gd I} \ldots s_{\gd J}$, $c_{\gd I \ldots J} = c_{\gd I} \ldots c_{\gd J}$, and similarly for $\gc$ instead of $\gd$. The coefficients for the mass and NUT charge are
\begin{align}
\label{munu}
\mu_1 & = 1 + \textstyle \sum_I [ \tfrac{1}{2} (s_{\gd I}^2 + s_{\gc I}^2) - s_{\gd I}^2 s_{\gc I}^2 ] + \tfrac{1}{2} \textstyle \sum_{I, J} s_{\gd I}^2 s_{\gc J}^2 , \nnr
\mu_2 & = \textstyle \sum_I s_{\gd I} c_{\gd I} [ (s_{\gc I} / c_{\gc I}) c_{\gc 1 2 3 4} - (c_{\gc I} / s_{\gc I}) s_{\gc 1 2 3 4} ] , \nnr
\nu_1 & = \textstyle \sum_I s_{\gc I} c_{\gc I} [ (c_{\gd I} / s_{\gd I}) s_{\gd 1 2 3 4} - (s_{\gd I} / c_{\gd I}) c_{\gd 1 2 3 4} ] , \nnr
\nu_2 & = \iota - D ,
\end{align}
where
\begin{align}
\iota & = c_{\gd 1 2 3 4}c_{\gc 1 2 3 4}+s_{\gd 1 2 3 4} s_{\gc 1 2 3 4} \nnr
& \quad + \textstyle \sum_{I < J} c_{\gd 1 2 3 4} (s_{\gd I J} / c_{\gd I J}) (c_{\gc I J} / s_{\gc I J}) s_{\gc 1 2 3 4} , \nnr
D & = c_{\gd 1 2 3 4}s_{\gc 1 2 3 4} + s_{\gd 1 2 3 4}c_{\gc 1 2 3 4} \nnr
& \quad + \textstyle \sum_{I < J} c_{\gd 1 2 3 4} (s_{\gd I J} / c_{\gd I J}) (s_{\gc I J} / c_{\gc I J}) c_{\gc 1 2 3 4} . \label{Dconstant}
\end{align}
The linear functions appearing in the solution are
\begin{align}
\label{linearfunctions}
L^I(r) & = (m \pi_2^I - n_0 \pi_1^I)r -4 (m^2+n_0^2)\frac{\p D}{\p \delta_I},\nnr
\widetilde{L}_I(r) & = (m \rho^2_I - n_0 \rho^1_I)r - 4 (m^2+n_0^2) \widetilde D_I,
\end{align}
where
\be
\wtd{D}_I = (s_{\gc I}/c_{\gc I}) c_{\gc 1 2 3 4} s_{\gd I}^2 - (c_{\gc I}/s_{\gc I}) s_{\gc 1 2 3 4} c_{\gd I}^2 .
\ee
The solution also involves the constants
\begin{align}
\label{Vconstants}
V^I & = (n_0 \rho^1_I -m \rho^2_I)n_0 + 2(m^2+n_0^2) \frac{\p C}{\p \delta_I} , \nnr
\widetilde V_I & = (n_0 \pi^I_1 - m \pi^I_2 )n_0 + 2(m^2+n_0^2) \widetilde C_I,
\end{align}
where
\begin{align}
C & = 1 + \textstyle  \sum_I (s_{\gd I}^2 c_{\gc I}^2 + s_{\gc I}^2 c_{\gd I}^2) + \textstyle  \sum_{I < J} (s_{\gd I J}^2 + s_{\gc I J}^2) \nnr
& \quad + \textstyle  \sum_{I \neq J} s_{\gd I}^2 s_{\gc J}^2 +  \textstyle \sum_I \sum_{J < K} (s_{\gd I}^2 s_{\gc J K}^2 + s_{\gc I}^2 s_{\gd J K}^2) \nnr
& \quad + 2 \textstyle  \sum_{I < J} [ s_{\gd 1 2 3 4} c_{\gd 1 2 3 4} (s_{\gc I J}/c_{\gd I J}) (c_{\gc I J}/s_{\gd I J}) \nnr
& \quad + s_{\gd 1 2 3 4}^2 (s_{\gc I J}^2 / s_{\gd I J}^2) + s_{\gd I J} s_{\gc I J} c_{\gd I J} c_{\gc I J} \nnr
& \quad + s_{\gd I J}^2 s_{\gc I J}^2 ] - \nu_1^2 - \nu_2^2 , \label{Cconstant}
\end{align}
and
\begin{align}
\wtd{C}_I & = (s_{\gd 1 2 3 4} - c_{\gd 1 2 3 4}) \wtd{C}_{I I} + 2 s_{\gc I} c_{\gc I} s_{\gd 1 2 3 4} ( 2 + \textstyle \! \sum_K s_{\gc K}^2 ) \nnr
& \quad + \textstyle \sum_{J \neq I} [ c_{\gd 1 2 3 4} s_{\gd I J} /c_{\gd I J} - s_{\gd 1 2 3 4} c_{\gd I J} /s_{\gd I J} ] \wtd{C}_{I J} \nnr
& \quad + 2 \textstyle \sum_{J \neq I} s_{\gc J} c_{\gc J} [ (s_{\gd I J}/c_{\gd I J}) c_{\gd 1 2 3 4} (s_{\gc I}^2 + s_{\gc J}^2) \nnr
& \quad - (c_{\gd I J}/s_{\gd I J}) s_{\gd 1 2 3 4} \textstyle \sum_{K \neq I, J} s_{\gc K}^2 ], \nnr
\wtd{C}_{I J} & = 2 (1 + 2 s_{\gd I}^2) s_{\gc 1 2 3 4} [ ( 2 + \textstyle \sum_{K \neq J} s_{\gc K}^{-2} ) s_{\gc 1 2 3 4} c_{\gc J}/s_{\gc J} \nnr
& \quad - (1 + 2 s_{\gc J}^2) c_{\gc 1 2 3 4}/(s_{\gc J} c_{\gc J}) ] \nnr
& \quad + 2 s_{\gd I}^2 s_{\gc J} c_{\gc J} ( 1 + \textstyle \sum_K s_{\gc K}^2 ) .
\end{align}
The coefficients for the $i = 1$ scalars are
\begin{align}
\label{xieta}
\xi_{11} & = [ (s_{\delta 123}c_{\delta 4} - c_{\delta 123}s_{\delta 4} )s_{\gamma_1}c_{\gamma_1} + (1 \leftrightarrow 4) ] \nnr
& \quad - ( (1,4) \leftrightarrow (2,3)) , \nnr
\xi_{12} & = [\tfrac{1}{2}(c_{\delta 23}s_{\gamma 14}+c_{\gamma 14} s_{\delta 23 })(c_{\delta 14}c_{\gamma 23} + s_{\gamma 23}s_{\delta 14}) \nnr
& \quad + s_{\gd 1} s_{\gc 4} c_{\gd 4} c_{\gc 1} (s_{\gd 2} s_{\gc 2} c_{\gd 3} c_{\gc 3} + s_{\gd 3} s_{\gc 3} c_{\gd 2} c_{\gc 2}) \nnr
& \quad + (1 \leftrightarrow 4)] - ( (1,4) \leftrightarrow (2,3)) , \nnr
\xi_{13} & = [( s_{\delta 1 3 4} c_{\delta 2} c_{\gamma 2}^2 + c_{\delta 1 3 4} s_{\delta 2} s_{\gamma 2}^2 ) s_{\gamma_3} c_{\gamma_3} + (2 \leftrightarrow 3) ] \nnr
& \quad - ( (1,4) \leftrightarrow (2,3)) , \nnr
\eta_{1 1} & = s_{\gd 2}^2 + s_{\gd 3}^2 + s_{\gc 1}^2 + s_{\gc 4}^2 + (s_{\gd 2}^2 + s_{\gd 3}^2) (s_{\gc 1}^2 + s_{\gc 4}^2) \nnr
& \quad + (s_{\gd 2}^2 - s_{\gd 3}^2) (s_{\gc 3}^2 - s_{\gc 2}^2) , \nnr
\eta_{1 2} & = 2 s_{\gd 2} c_{\gd 2} (c_{\gc 2} s_{\gc 1 3 4} - s_{\gc 2} c_{\gc 1 3 4}) + (2 \leftrightarrow 3) , \nnr
\eta_{1 3} & = 2 s_{\gd 2 3} c_{\gd 2 3} (s_{\gc 2 3} c_{\gc 2 3} + s_{\gc 1 4} c_{\gc 1 4})+ s_{\gd 2 3}^2 (1 + \textstyle \sum_I s_{\gc I}^2 ) \nnr
& \quad + (s_{\gd 2}^2 + s_{\gd 3}^2+2 s_{\delta 23}^2) (s_{\gc 1 4}^2 + s_{\gc 2 3}^2) \nnr
& \quad + s_{\gd 2}^2 s_{\gc 2}^2 + s_{\gd 3}^2 s_{\gc 3}^2 + s_{\gc 1 4}^2 .
\end{align}
The results for $i = 2$ and $i = 3$ are obtained by respectively interchanging indices $1 \leftrightarrow 2$ and $1 \leftrightarrow 3$.

\end{document}